\documentclass[useAMS,usenatbib,usegraphicx]{mn2e}

\usepackage{lscape} 

\title[SINFONI/VLT 3D spectroscopy of massive galaxies: Evidence of rotational support at $z\sim1.4$]{Kinematics of massive galaxies: evidence of rotational support at z $\sim$ 1.4}
\author[F. Buitrago et al.]{Fernando Buitrago$^{1,2}$ \thanks{E-mail: fb@roe.ac.uk}\thanks{Based on observations made with ESO telescopes at the La Silla Paranal Observatory under programme ID 079.B-0430(A)}, 
Christopher J. Conselice$^{1}$, Beno\^{i}t Epinat$^{3,4,5}$, \newauthor Alejandro G. Bedregal$^{6,1,7}$, Ruth Gr\"{u}tzbauch$^{8}$, Benjamin J. Weiner$^{9}$ \\
\\
\\
$^{1}$University of Nottingham, School of Physics \& Astronomy, Nottingham, NG7 2RD U.K. \\
$^{2}$SUPA\thanks{Scottish Universities Physics Alliance}, Institute for Astronomy, University of Edinburgh, Royal Observatory, Edinburgh, EH9 3HJ, U.K. \\
$^{3}$Universit\'e de Toulouse; UPS-OMP; IRAP; F-31400 Toulouse, France \\
$^{4}$CNRS; IRAP; 14, avenue Edouard Belin, F-31400 Toulouse, France \\
$^{5}$Aix Marseille Universit\'e, CNRS, LAM (Laboratoire d'Astrophysique de Marseille) UMR 7326, 13388, Marseille, France \\
$^{6}$Centro de Astrobiolog\'ia, INTA-CSIC, E-28850 Torrej\'on de Ardoz, Madrid, Spain \\
$^{7}$University of Minnesota, 116 Church Street SE, Minneapolis, MN 55455, USA \\
$^{8}$Center for Astronomy and Astrophysics, Observatorio Astronomico de Lisboa, Tapada da Ajuda, 1349-018 Lisboa, Portugal \\
$^{9}$Steward Observatory, 933 N. Cherry St., University of Arizona, Tucson, AZ 85721, USA }

\def\halpha{H$\alpha$\,}
\def\ks{K$_{s}$\,}
\begin{document}


\maketitle

\label{firstpage}

\begin{abstract}
There is cumulative evidence showing that, for the most massive galaxies, the
fraction of disk-like objects compared to those with spheroidal properties
increases with redshift. However, this evolution is thus far based on the
surface brightness study of these objects. To explore the consistency of this scenario, it
is necessary to measure the dynamical status of these galaxies. With this
aim we have obtained seeing-limited near-infrared integral field spectra in the H-band for 10 massive galaxies
(M$_{*}\geq10^{11} h_{70}^{-2}$ M$_{\odot}$) at $z\sim1.4$  with SINFONI at the VLT. 
Our sample is selected by their stellar mass and EW[OII]$> 15$ \AA, to secure
their kinematic measurements, but without accounting for any morphological or flux criteria a priori.
Through this 3D kinematic spectroscopy
analysis we find that half (i.e. 50$\pm$7\%) of our  galaxies are compatible
with being rotationally supported disks, in agreement with previous morphological
expectations. This is a factor of approximately two higher than what is observed in the
present Universe for objects of the same stellar mass. Strikingly, the majority of our sample of
massive galaxies show extended and fairly high rotational velocity maps, implying that massive
galaxies acquire rapidly rotational support and hence gravitational equilibrium.  Our sample
also show evidence for ongoing interactions and mergers. Summarizing, massive galaxies
at high-z show a significant diversity
and must have continued evolution beyond the fading of stellar populations, to become
their present day counterparts.

\end{abstract}



\section{Introduction}
\label{sec:intro}

Massive (M$_{stellar}\geq10^{11} h_{70}^{-2}$ M$_{\odot}$) galaxies represent a challenge to the dominant  $\Lambda CDM$ paradigm as their properties (such as number densities, star formation histories, size
 growth, etc.) should be well constrained by galaxy evolution models, but seemingly this is not always the case (e.g., Benson et al. 2003, Baugh et al. 2006; Conselice et al. 2007, Trujillo 2012). Several
 recent studies find that these galaxies are better represented by a disk-like population at z$  = $2-3 (Cameron et al. 2011, Van der Wel et al. 2011, Bruce et al. 2012, Buitrago et al. 2013) as opposed to
 their counterparts in the local Universe, which are preferentially large early-type galaxies (Baldry et al. 2004). A dynamical test of this scenario is still missing, and would complement the great deal of
 morphological information gathered to date (Oesch et al. 2010, Weinzirl et al. 2011, Wang et al. 2012, Van der Wel et al. 2013 in prep.).

Observationally, most of the information regarding high redshift galaxies comes from large and deep NIR surveys, which aim to probe galaxy evolution during the last 12 Gyr. The data gathered so far favours
 a picture in which late-type and clumpy/interacting objects are more common at high redshift, due to the higher gas fractions for these systems (Elmegreen \& Elmegreen 2006, Erb et al. 2006, Bournaud et al. 2008, Tacconi et al. 2010) and higher merging
 rates (e.g. Conselice et al. 2008, L\'{o}pez-Sanjuan et al. 2011, Bluck et al. 2012). Ultimately, we would ideally like to rely on spectroscopic information to fully test and characterize all the processes
 involved in galaxy assembly. However, even for massive (and thus very often luminous) galaxies, it is very expensive in observational time to obtain high signal-to-noise ratio spectra. From the tens of
 massive galaxies studied so far at $z > 1.5$ with traditional long-slit techniques (Kriek et al. 2006, Cimatti et al. 2008, Newman et al. 2010, Onodera et al. 2010, van de Sande et al. 2011, Toft et al.
 2012) there is some agreement on the high velocity dispersion values of these objects, confirming their inherent massive nature even at such early cosmic times (Cenarro \& Trujillo 2009, Cappellari et al.
 2009, Van de Sande et al. 2012). Despite this, the true nature of these systems and how they evolve is not yet understood.

Integral field spectroscopy (aka 3D spectroscopy) is presently a well-established technique which can enhance greatly our understanding of these massive galaxies, both at low-redshift (Cappellari et al.
 2011, S\'{a}nchez et al. 2012) and at high-redshift (Law et al. 2009, F\"orster-Schreiber et al. 2009, Contini et al. 2012). On the one hand, it measures the rotational and the velocity dispersion support
 (currently from gaseous kinematics) for a galaxy, and thus provides us with physical information, such as both baryonic and dark matter components, as well as the presence of rotation. Spatially
-distributed spectral information furthermore helps us address the question of how morphology and galaxy assembly are linked (Weiner et al. 2006, Ceverino et al. 2012, L\'opez-Sanjuan et al. 2013). State-of-the-art Integral Field Unit (IFU) studies reveal a kinematic mixture
 within high-z systems (e.g., Genzel et al. 2008, Nesvadba et al. 2008, van Starkenburgh et al. 2008, Shapiro et al. 2008, Bournaud et al. 2008, Cresci et al. 2009, Law et al. 2007, 2009, Wright et al. 2007, 2009, Epinat et al. 2009, Bouch\'{e} et al. 2010, Lemoine-Busserolle et al. 2010,
 Lemoine-Busserolle \& Lamareille 2010, Gnerucci et al. 2011, Epinat et al. 2012), with a high percentage of objects displaying large ordered rotational motions, and also fairly large velocity dispersions.

In the low redshift Universe, the SAURON survey (de Zeeuw et al. 2002) and the more recent ATLAS$^{3D}$ survey (Cappellari et al. 2011) have opened a new perspective on the kinematics of local massive
 galaxies. They classify early-type galaxies as slow and fast rotators depending on the degree of angular momentum they exhibit. Fast rotators have been found to host disks made up of gas and stars which
 contain a range of the galactic mass fraction (Falc\'on-Barroso et al. 2006, Krajnovi\'c et al. 2008, Davis et al. 2011, Krajnovi\'c et al. 2012). In the other hand, at high-z, not only massive galaxies
 seem to be better represented by a disk-like population (e.g., Buitrago et al. 2013) but, for those with early-type morphology, a flatter nature (possibly indicating a disk-component) is prevalent (Van der
 Wel et al. 2011, Chevance et al. 2012, Chang et al. 2013a, Chang et al. 2013b in prep.). Our study aims to provide data on the kinematics of massive galaxies at intermediate redshifts (z $\sim$ 1.4) and
 thus covering the gap between current and distant cosmic epochs. 

We discuss in this paper IFU studies of a sample of galaxies with log M$_{*} / $M$_{\odot} >$ 11 at high (z $\sim$ 1.4) redshift. These are good targets for IFU spectroscopy due to their relative compactness (which make
 them easy to observe even with a small field-of-view) and the current lack of spectra for this galaxy population. Ideally, absorption lines measurements would be the best indicators to examine the stellar
 populations and the motions within this population  (e. g. Bedregal et al. 2009). Nevertheless, \halpha emission line analyses are usually preferred due to its relative ease of study. It is very important
 to emphasize the fact that one must be cautious when interpreting the derived \halpha kinematics, since this ionized gas is collisional and dissipative and it may not be coupled with the stellar component
 in the galaxy. However, it is not unreasonable to expect good correlation between gas emission and broadband imaging (e.g. F\"orster-Schreiber et al. 2011), especially when studying relaxed systems.  


We present in this paper 8m-telescope VLT observations of these massive galaxies using the SINFONI IFU spectroscopy instrument. Our sample comprises 10 objects at $z\sim1.4$, whose redshift choice is a
 trade-off between high redshift and having a sample selected by stellar mass. Nevertheless, this redshift is key on the development of massive galaxies in particular, and galaxies in
 general, as it nearly coincides with the star formation and quasar activity peak, apart from being the epoch at which massive galaxies begin to switch their morphologies from late to early types (Buitrago
 et al. 2013). We discuss the derived H$\alpha$ kinematic properties for this sample and interpret these findings in the context of massive galaxy formation. 

This paper is structured as follows: Section \ref{sec:data} describes the data reduction and subsequent analysis, Section \ref{sec:results} discusses the different tests 3D spectroscopy offer for
 characterizing the rotation dominated nature of these objects, and in Section \ref{sec:conclusions} we present the conclusions of our study. There is a final Appendix in which we show each galaxy,
 explaining in a detailed way its particularities, and how every massive galaxy is related with the total sample. We name the galaxies in our sample with the prefix POWIR followed by a number, instead of
 the numeric code in the parent survey, as it simplifies tracking the individual galaxies throughout the paper (please note there is not a POWIR9 galaxy). Besides, these numbers are plotted along with the
 galaxy symbols aiming at the same purpose. We assume a concordant cosmology: H$_{0}$=70 km s$^{-1}$Mpc$^{-1}$, $\Omega_{\lambda}$ = 0.7, and $\Omega_{\rm m}$ = 0.3, and use a Chabrier (2003) IMF and AB
 magnitude units unless otherwise stated. 
%

\section{Data and analysis}
\label{sec:data}

\subsection{Observations}
\label{subsec:observations}

\begin{table*}
\begin{minipage}{\textwidth}
\caption{Observational data for our massive galaxy sample}
\label{tab:structural_data} 
\resizebox{\textwidth}{!}{
\begin{tabular}{c|c|c|c|c|c|c|c|c|c|c|c|c|c}
Name & POWIR ID & RA        & DEC        & z   & log M$_{*}$                     & $K_{s}$ mag     & EW$_{[OII]}$ & log L$_{[OII]}$  & SFR                   & Observ. night & Integration time & (S/N)$_{threshold}$  & Seeing \\
     &          & (J2000)   & (J2000)    &     & log ($h_{70}^{-2}$ M$_{\odot}$) & Vega magnitudes & \AA          & log erg s$^{-1}$ & M$_{\odot}$ yr$^{-1}$ &               &                  & sec                  & arcsec \\
(1)  & (2)      & (3)       & (4)        & (5) & (6)                             & (7)             & (8)          & (9)              & (10)                  & (11)          & (12)             & (13)                 & (14) \\
    \hline
    POWIR1  & $ 32007614 $ & $ 23.516207 $ & $ 0.043184 $ & $ 1.374 $ & $ 11.2 $ & $ 18.2 $ & $32.9\pm2.2$  & $42.21\pm0.03$ & $17.78$ & $ 21-Jun $ & $ 5400 $ & $ 3 $ & $ 0.58 $ \\
    POWIR2  & $ 32013051 $ & $ 23.519257 $ & $ 0.111119 $ & $ 1.396 $ & $ 11.0 $ & $ 18.6 $ & $17.7\pm7.5$  & $41.94\pm0.18$ & $11.51$ & $ 04-Sep $ & $ 5400 $ & $ 3 $ & $ 0.52 $ \\
    POWIR3  & $ 32015443 $ & $ 23.500269 $ & $ 0.146335 $ & $ 1.384 $ & $ 11.2 $ & $ 18.5 $ & $21.5\pm9.1$  & $42.11\pm0.18$ & $15.13$ & $ 08-Aug $ & $ 2700 $ & $ 3 $ & $ 0.65 $ \\
    POWIR4  & $ 32015501 $ & $ 23.501966 $ & $ 0.102931 $ & $ 1.394 $ & $ 11.4 $ & $ 18.0 $ & $15.1\pm1.8$  & $42.37\pm0.05$ & $23.01$ & $ 31-Jul $ & $ 2700 $ & $ 2 $ & $ 0.58 $ \\
    POWIR5  & $ 32021317 $ & $ 23.499363 $ & $ 0.161825 $ & $ 1.382 $ & $ 11.3 $ & $ 18.3 $ & $15.1\pm1.7$  & $41.92\pm0.05$ & $11.14$ & $ 20-Jul $ & $ 5400 $ & $ 2 $ & $ 0.55 $ \\
    POWIR6  & $ 32021394 $ & $ 23.496317 $ & $ 0.159318 $ & $ 1.375 $ & $ 11.5 $ & $ 17.9 $ & $12.4\pm1.8$  & $41.87\pm0.06$ & $10.28$ & $ 01-Jul $ & $ 5400 $ & $ 3 $ & $ 0.48 $ \\
    POWIR7  & $ 32029850 $ & $ 23.517859 $ & $ 0.282879 $ & $ 1.396 $ & $ 11.3 $ & $ 18.3 $ & $32.1\pm32.1$ & $42.13\pm1.87$ & $15.63$ & $ 19-Jul $ & $ 5400 $ & $ 2 $ & $ 0.63 $ \\
    POWIR8  & $ 32037003 $ & $ 23.503817 $ & $ 0.334799 $ & $ 1.400 $ & $ 11.0 $ & $ 18.4 $ & $72.8\pm72.8$ & $42.85\pm7.70$ & $49.88$ & $ 01-Jul $ & $ 5400 $ & $ 3 $ & $ 0.42 $ \\
    POWIR9  & $ -        $ & $ -         $ & $ -        $ & $ -     $ & $ -    $ & $ -    $ & $ - $         & $ - $          & $ - $   & $ - $      & $ -    $ & $ - $ & $ - $    \\
    POWIR10 & $ 32002481 $ & $ 23.516451 $ & $ 0.024167 $ & $ 1.389 $ & $ 11.1 $ & $ 18.7 $ & $25.2\pm5.4$  & $42.01\pm0.09$ & $12.88$ & $ 20-Jul $ & $ 5400 $ & $ 3 $ & $ 0.55 $ \\
    POWIR11 & $ 32100778 $ & $ 23.510761 $ & $ 0.235435 $ & $ 1.393 $ & $ 11.1 $ & $ 18.3 $ & $21.9\pm2.6$  & $42.28\pm0.05$ & $19.90$ & $ 08-Aug $ & $ 2700 $ & $ 2 $ & $ 0.65 $ \\
\hline
\end{tabular}
}\\
Notes. (1) Name of the galaxy (2) Name of the galaxy in the parent POWIR/DEEP2 survey (3) Right ascension (4) Declination (5) Spectroscopic redshift from our SINFONI observations (6) Stellar mass from the
 parent POWIR/DEEP2 survey (7) $K_{s}$-band magnitude from the parent POWIR/DEEP2 survey (8) [OII] Equivalent Width (9) [OII] Luminosity (10) Star Formation Rate from the [OII] Luminosity
 (11) Date of the observations, in 2007 (12) Integration time (13) Signal-to-noise ratio threshold above which we show
 the spaxels in the kinematic maps (14) Seeing as derived from the SINFONI telluric standards observed.
\end{minipage}
\end{table*}

The parent sample from which our target galaxies are selected is the Palomar Observatory Wide InfraRed survey (POWIR; Bundy et al. 2006, Conselice et al. 2007, 2008).   This survey covers a $1.53$
 deg$^{2}$ area in the \ks and J bands down to $K_{Vega} = 21$ and $J_{Vega} = 23.5$. This imaging consists of 75 Palomar WIRC camera pointings, with a pixel scale of 0.25$"/$pix. In the \ks band 30-s
 exposures were taken, for total 1-2h integration time per pointing, and with typical seeing of $0.7-1"$. In addition, optical coverage was supplied with the 3.6m Canada-France-Hawaii Telescope (CFHT) using
 the CFH12K camera in the B, R and I bands. The R-band depth is $R_{AB} = 25.1$, with similar results for the other two bands (see Coil et al. 2004 for more details). Both Palomar and CFHT images were
 analysed using $2"$ diameter apertures. 

Stellar masses were derived with the photometric techniques discussed in Bundy et al. (2006) using a Chabrier (2003) IMF. Our stellar mass computational method consists of fitting a grid of model Spectral Energy Distributions (SEDs)
 constructed from Bruzual \& Charlot (2003) stellar populations synthesis models. The Star Formation histories are parametrized by utilizing so-called tau-models (SFR $\propto \rm exp(-t/\tau)$; with $\tau$
 randomly selected from a range between 0.01 and 10 Gyr, and the age of the onset of star formation ranging from 0 to 10 Gyr), with a range of metallicities (from 0.0001 to 0.05) and dust contents
 (parametrized by the effective V-band optical depth $\tau_{V}$, where the used values were $\tau_{V}$ = 0.0, 0.5, 1, 2). To analyze the impact of Thermally-Pulsating AGB stars, the same exercise was also performed with
 Charlot \& Bruzual (2007) models, inferring slightly smaller masses (on the order of $\sim10\%$). It turns out that parameters such as metallicity, e-folding time or age are not as well constrained as
 stellar mass due to the various degeneracies. The final error in stellar mass is measured as 0.2-0.3 dex, i.e. roughly a factor of two (Bundy et al. 2006, Conselice et al. 2007, Gr\"utzbauch et al. 2011).

\begin{figure}
\includegraphics[angle=0,width=1.0\linewidth]{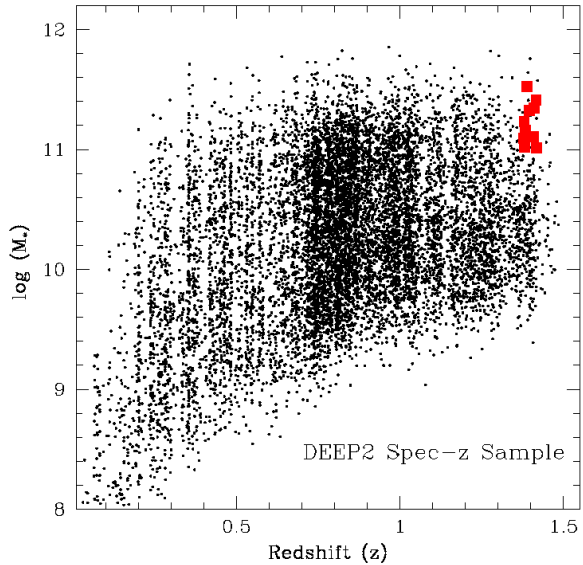}

\caption{Mass-redshift relation for the DEEP2 spectroscopic sample. Our massive galaxies are highlighted by the red squares. Our study is an attempt to characterise the high redshift high mass population of this spectroscopic survey.}
\label{fig:deep2}

\end{figure}

Spectroscopic redshifts were measured within the DEEP2 Redshift Survey\footnote{http://deep.ps.uci.edu/dr4/credit.html} (Davis et al. 2003, 2007, Newman et al. 2013) using the DEIMOS spectrograph (Faber et al. 2003) at the Keck II telescope. Spectra were obtained with a resolution
 of $R\sim5000$ within the wavelength range $6500-9100$ \AA. Redshifts were measured comparing templates to the data, and we utilised only those in which two or more lines were identified. We show in Figure
 \ref{fig:deep2} how our massive galaxy sample compares to the rest of the DEEP2 spectroscopic sample. It is noteworthy that our final choice of galaxies to observe was not based on a selection using
 colours, morphologies or sub-mm flux as many other IFU studies. Additionally, we have not only explored compact massive galaxies. Instead our sample is solely selected by stellar mass without any other
 criteria a priori except with an [O II] Equivalent Width (EW$_{[OII]}$) $> 15$ \AA. This last condition might bias our sample towards investigating  star forming systems. However, this is a necessary requirement to
 robustly asses our galaxy sample's kinematic features. 

\begin{figure}
\includegraphics[angle=0,width=1.0\linewidth]{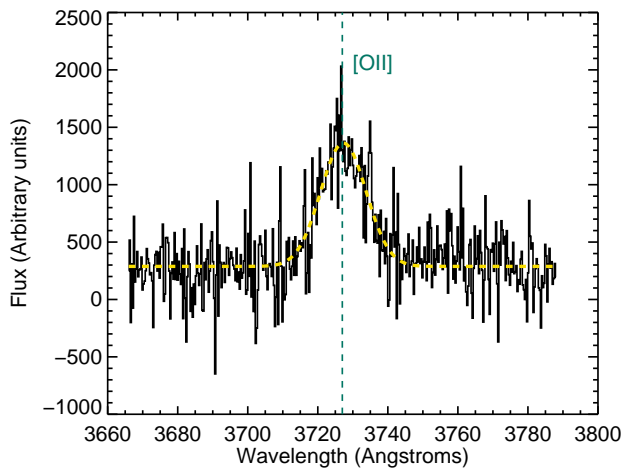}

\caption{Stacked DEEP2 spectra for the massive galaxies in our sample around the [OII]$\lambda$3727 emission line wavelength. 
Detecting this emission line in these massive galaxies at z $\sim$ 1.4 indicates they are not devoid of star formation (Weiner et al. 2007, Noeske et al. 2007) 
and makes them interesting candidates for 3D spectroscopy investigations given their high stellar mass.}
\label{fig:O_II_stack}

\end{figure}

Figure \ref{fig:O_II_stack} shows the stacked DEEP2 spectra around the [OII] emission line for our galaxy sample.
The existence of this line demonstrates (in conjunction with Fig. \ref{fig:deep2}) that a high fraction of the 
massive objects at z $\sim$1.4 are not devoid of star formation, as
 highlighted by various studies for massive galaxies (e.g., Conselice et al. 2007, P\'{e}rez-Gonzalez et al. 2008, Cava et al. 2010, Bauer et al. 2011, Viero et al. 2012).
Adding this information along with the fact that various morphologies are found in this work
 suggests that we are close to having a representative sample selected by stellar mass. 

The most important parameters for our galaxy sample are listed in Table \ref{tab:structural_data},
along with the individual EWs, and a rough estimation of the Star Formation Rate (SFR) for each massive galaxy.
This later value comes from the prescription found in Weiner et al. (2007) and Noeske et al. (2007) (see also Kewley, Geller \& Jansen 2004):
\[
log SFR = log(0.7) + logL_{OII} - log\frac{[OII]}{H\alpha} + logA(H\alpha) - log L_{H\alpha}
\]
where we took $[OII]/H\alpha$ = 0.69, $A(H\alpha)$ = 3.3 and $log L_{H\alpha}$ =  41.104, corrected to our Chabrier IMF by means of the 0.7 factor.
The $[OII]/H\alpha$ and extinction ratios are from average properties of lower-z emission line galaxies in DEEP2, while the factor to convert L$_{H\alpha}$
into SFR comes from a standard Kennicutt SFR prescription (Kennicutt, Tamblyn \& Congdon 1994; Kennicutt 1998).

Our group was granted 20 hours of observing time in service mode with the 3D-spectrograph SINFONI (Eisenhauer et al. 2003, Bonnet al. 2004) at ESO-VLT located in UT4-Yepun. Our observations were conducted
 during 9 nights from June to September 2007 -- ESO run ID 079.B-0430(A) --. SINFONI was used in seeing limited mode and thus with a spatial sampling of $0.125"\times0.25"$. Due to the redshifts of our
 sample of galaxies we chose to observe them in the H-band in order to map the \halpha emission. We also took special care in our final galaxy selection to ensure that their \halpha emission lines were not
 situated close to any OH sky emission lines -- based on the atlas from Rousselot et al. (2000) -- which would potentially hamper our results. The spectral resolution ($R\sim3000$) allows us to disentangle
 sky emission lines close to our target. 

Our observational strategy was the so-called `butterfly pattern' or `on-source dithering', by which the galaxy is set in two opposite corners of the detector to remove sky background using contiguous frames
 in time. Several galaxies in our sample (POWIR4, POWIR5 and POWIR7) could only be observed half of their nominal integration time (1h 30min).  Even in these cases, exquisite SINFONI sensitivity permitted
 us to detect the emission from all our objects. Images were dithered by $0.3"$ in order to minimize instrumental artefacts when the individual observations were aligned and combined together. PSF and
 telluric stars were also observed along with each galaxy for calibration purposes. Measured PSFs are listed in Table \ref{tab:structural_data}, for a mean seeing of $0.56 \arcsec$ throughout our
 observations.

\subsection{Data reduction \& observed kinematic maps}
\label{subsec:data_reduction}

We have used the ESO-SINFONI pipeline version 2.5.0 (Modigliani et al. 2007, Mirny et al. 2010) to reduce our data. In brief, this pipeline subtracts sky emission lines (using algorithms by Davies et al.
 2007), corrects the image using darks and flat-fields, spectrally calibrates each individual observation and reconstructs all the information into a final datacube. The recipe used for this purpose was
 \texttt{sinfo\_rec\_jitter}, which was fed exclusively with the master files provided by ESO. All of these processes were performed separately for each individual exposure. Afterwards the two datacubes
 were combined into a single one by using the recipe \texttt{sinfo\_utl\_cube\_combine}.   We always used the pipeline parameter \textit{product-density = 3} (which retrieves the most detailed possible
 outputs), \textit{objnod-scales\_sky = true} (to perform a subtraction of the median value at each wavelength and thus remove the sky more efficiently) and \textit{skycor.rot\_cor = true} (to remove the
 contribution of any rotational OH transitions).

The final datacube was spatially smoothed using a sub-seeing Gaussian core (FWHM=2 pixels) to increase the Signal-to-Noise Ratio (SNR) without affecting our data interpretation.   We analysed this  datacube
 with IDL routines we constructed. Basically, we located the \halpha line in each spaxel according to the known spectroscopic redshift of the target galaxy, and then fit a Gaussian profile, taking into
 account the sky spectrum weighting its contribution with the help of the routines \texttt{mpfit} and \texttt{mpfitfun} (Markwardt et al. 2009). Radial velocity maps were computed using the relativistic
 velocity addition law:

$$V_{spaxel} = \frac{(z_{spaxel}-z_{cen})}{1+z_{cen}}c$$

\noindent where $z_{spaxel}$ and $z_{cen}$ are the redshifts for a given spaxel and for the kinematic centre of the galaxy, respectively. 

From the \halpha line width, we computed velocity dispersion maps, subtracting the instrumental broadening, measured from sky lines. In addition, we obtained  \halpha and [NII] $\lambda6583\rm \AA$  line
 flux maps. The SNR per spaxel was calculated in the following manner: the signal was the intensity of the \halpha line, and the noise was the standard deviation of the residual spectrum, with both signal
 and noise weighted by the sky contribution around the \halpha wavelength. A continuum map was also constructed with the spectral information in the range $1.5 - 1.7 \mu m$, i.e., in all the H-band except
 its borders, where the information was noisier. We fit a linear function to the galaxy spectrum in this wavelength range in order to account for the existence of continuum emission and its possible
 variation within this wavelength range. Our continuum maps show the integral of the fitted mathematical function. Spaxels with values below zero are coloured in white. 

All our objects have continuum emission which we have compared to the ionized gas emission, which in principle tells us only about the areas of star formation in each galaxy. As such, comparing the peak of
 the emission and the continuum gives us insights to how well \halpha, and hence star formation, traces the underlying older stellar populations. One caveat to this was a known problem with the SINFONI
 detector, where there are stripes of flux in the data after coadding high numbers of spectral pixels. One can see these stripes for instance in the continuum maps of POWIR1 (see Appendix A Fig. \ref{fig:montage_powir1}; southern part
 of the galaxy) and POWIR3 (see Appendix A Fig. \ref{fig:montage_powir3}; white stripe on the top of the galaxy). This effect prevented us from making a total continuum flux measurement, but did not affect qualitatively the fact that we
 could locate where the maximum of the continuum was in the detector.

\subsection{Data modelling}
\label{subsec:modelling}

We recovered the kinematic parameters for each galaxy in our sample by fitting a model to the velocity field obtained from our SINFONI datacubes. To perform this task we assumed that the galaxies from our
 sample are described kinematically as rotating disk systems with a symmetric rotation curve.  For this method we utilised the formalism and programs developed in Epinat et al. (2009).  The full theoretical
 description of this method is in Epinat et al. (2010), where the authors also conducted a comparison with local galaxies to asses the reliability of their method.  Essentially this consists of a $\chi^{2}$
 minimization between the observed data and a given high resolution model convolved to our pixel scale and seeing conditions. We chose the flat rotation curve parametrization used by Wright et al. (2007,
 2009), as suggested in Epinat et al. (2010) from the study of local galaxy velocity fields projected at high redshift. The analytical expression for this is given by:
$$V(r) = V_{t}\frac{r}{r_{t}}$$
when $r \leq r_{t}$ and 
$$V(r) = V_{t}$$ 
otherwise.  In the above equations $V_{t}$ is the value for the plateau in the rotation curve and $r_{t}$ is the radius at which the plateau is reached. The model contains seven parameters: the center 
($x_{c}$ and $y_{c}$), the systemic redshift (or velocity), the inclination of the disk, the position angle of the major axis and the two rotation curve parameters. Note that the fit to these simple
 formulae were done by considering the associated error map for the velocity field.

As shown and discussed in Epinat et al. (2010), due to the reduced spatial information of our data, and due to some degeneracy in the models, the center and the inclination are the least constrained
 parameters. We thus fixed the center to the spaxel with the maximum flux in the continuum maps (in principle the continuum may trace better the galactic center) as well as the inclination, reducing to four
 the number of free parameters in our model. In rotating disk models there is also a degeneracy between rotation velocity and inclination (its sine) that can only be solved using very high resolution data.
 As a result the inclination is the major source of uncertainty for determining the actual rotation velocity. Given the photometric quality of our POWIR parent sample imaging, it is difficult to constrain
 this parameter with a high certainty. We used GALFIT (Peng et al. 2010) to fit single S\'ersic (1968) surface brightness profiles to our sample. We have only utilised the inclination retrieved by the
 GALFIT analysis, and discarded the rest of the structural parameters, owing to the high uncertainties given the large Point Spread Function (PSF) of our images. We used bright, non-saturated stars within
 our imaging as a model PSF. The output inclination was then utilised as input for our velocity modelling. 

Once the best model for the rotational velocity was obtained, we also computed a model velocity dispersion map. To calculate this we took into account the width of the \halpha line due to the unresolved
 velocity gradient. The intrinsic velocity dispersion is obtained after subtracting in quadrature the velocity dispersion model map from the observed one. Finally, to facilitate a comparison with other
 samples and to discuss each galaxy as a whole, we also computed the velocity dispersion value doing a weighted average of the value of every spaxel by an amount inversely proportional to their squared
 error. Note also that this velocity dispersion is computed from the beam smearing corrected maps. To avoid confusion, henceforth we will refer to this quantity as the $1/$error$^{2}$ velocity dispersion.
 Results from the models are listed in Table \ref{tab:velocities}, and their maps are the first two montages in the Appendix A. All of the galaxy kinematic maps are also in the appendix. 

\section{Results}
\label{sec:results}

\subsection{Kinematic classification}
\label{subsec:kin_class}

Previous 3D spectroscopy studies of high redshift galaxies have demonstrated that these systems are more clumpy/irregular and have higher velocity dispersions than local galaxies (F\"orster-Schreiber et al.
 2011 and references therein). There are several attempts in the literature to establish a kinematic classification of high-z galaxies relying on \halpha kinematics (Flores et al. 2006, Law et al. 2009,
 Cresci et al. 2009, F\"orster-Schreiber et al. 2009, Epinat et al. 2009, Lemoine-Busserolle et al. 2010, Gnerucci et al. 2011, Epinat et al. 2012). All of these studies roughly agree that there are three basic kinematic classes, which
 may be linked with the morphological nature of each galaxy. First, rotating disks have been observed, showing well-defined and regular rotational velocity gradients that are larger than their velocity
 dispersion. Usually these systems are large in size. Ongoing mergers are also clearly distinguished, not only by disentangling two or more components but through a chaotic velocity pattern, and local
 increments in the velocity dispersion. Finally, objects which do not fit in any of the previous categories are tagged as perturbed rotators, which are probably more similar to early type systems due to
 their high velocity dispersion in comparison with their maximum rotational velocity.

Before characterising our sample according to these criteria, we enumerate several caveats that might affect our interpretation of the data. First, behind this classification there is the disk-like
 assumption in the modelling. This will not be an accurate model when dealing with mergers or pure spheroidal galaxies. Arguably, this has an impact in our sample as massive galaxies in the local Universe
 are predominantly early types (e. g. Baldry et al. 2004). However, at the redshift of our observations ($z \sim 1.4$), we would expect to have a morphological mixture (van Dokkum et al. 2011, Buitrago et
 al. 2013). Finally, we must remember we are looking at the gas emission and not at the total stellar component. 

Performing a preliminary investigation, we notice that the \halpha emission extends all over the \ks and continuum images for most cases, which is difficult to reconcile with the possibility of being
 spheroid-like objects, especially when \halpha emission is usually linked with star formation. As stated in the introduction, the existence of gas disks within early type galaxies has been reported in the
 past (Falc\'on-Barroso et al. 2006, Krajnovi\'c et al. 2008, Oosterloo et al. 2010, Krajnovi\'c et al. 2013) but their sizes (hundreds of parsecs) are much smaller than our current gas disks which span the
 galactic size. In fact, these disks are a fundamental aspect for understanding the evolution of these galaxies.

We focus now on the individual properties of the sample, which has been discussed at length in the Appendix A. In summary, the velocity fields of POWIR1, 2, 6 and 8 show clear gradients that are compatible with rotation. However, the centers of the
 latter two display fairly large velocity dispersions. In both Epinat et al. (2009) and F\"orster-Schreiber et al. (2009) these were indications of disks galaxies developing a bulge component. It is also
 noteworthy to mention that, in case of a very steep gradient or ring in \halpha, a central peak due to beam smearing is expected. However, these gas distributions are mostly found in the local Universe in
 galaxies with large bulges (Peletier et al. 2007), and thus this could be in agreement with a progressive evolution towards early-type systems. Nevertheless, in light of the kinematic probes, we classify
 the four galaxies as rotating disks. 

\begin{figure*}
\includegraphics[angle=0,width=1.\linewidth]{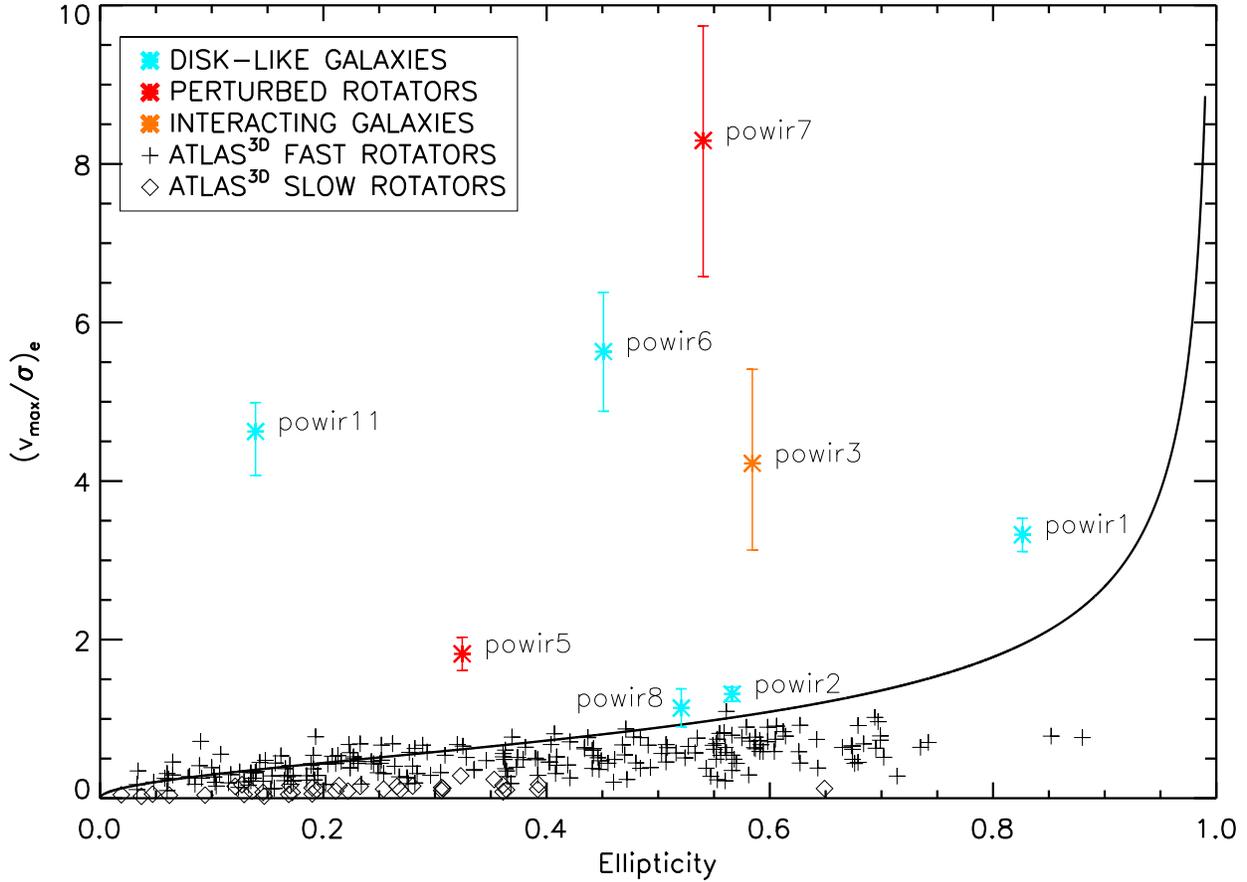}

\caption{($V_{max}/\sigma$,$\varepsilon$) diagram for massive ($M_{stellar}\geq10^{11} h_{70}^{-2} M_{\odot}$) galaxies in our sample, also called the anisotropy plot. Apart from these, we supplemented the
 figure with the local early-type galaxies in Emsellem et al. (2011). These low redshift objects are early-type galaxies studied as part of the ATLAS$^{3D}$ survey (Cappellari et al. 2011). Ellipticities
 for our sample were measured in the K-band imaging of POWIR/DEEP2 survey using GALFIT and thus taking into account the PSF of our imaging. The continuous line defines the ideal oblate rotator with
 isotropic stellar velocity distribution for integral field studies (Binney et al. 2005, Cappellari et al. 2007). The uncertainties are large, but it is clear that massive galaxies at $z\geq1.4$ depart from
 the low-z velocity dispersion dominated objects.}
\label{fig:anisotropy_plot}

\end{figure*}

\begin{figure*}
\includegraphics[angle=0,width=1.\linewidth]{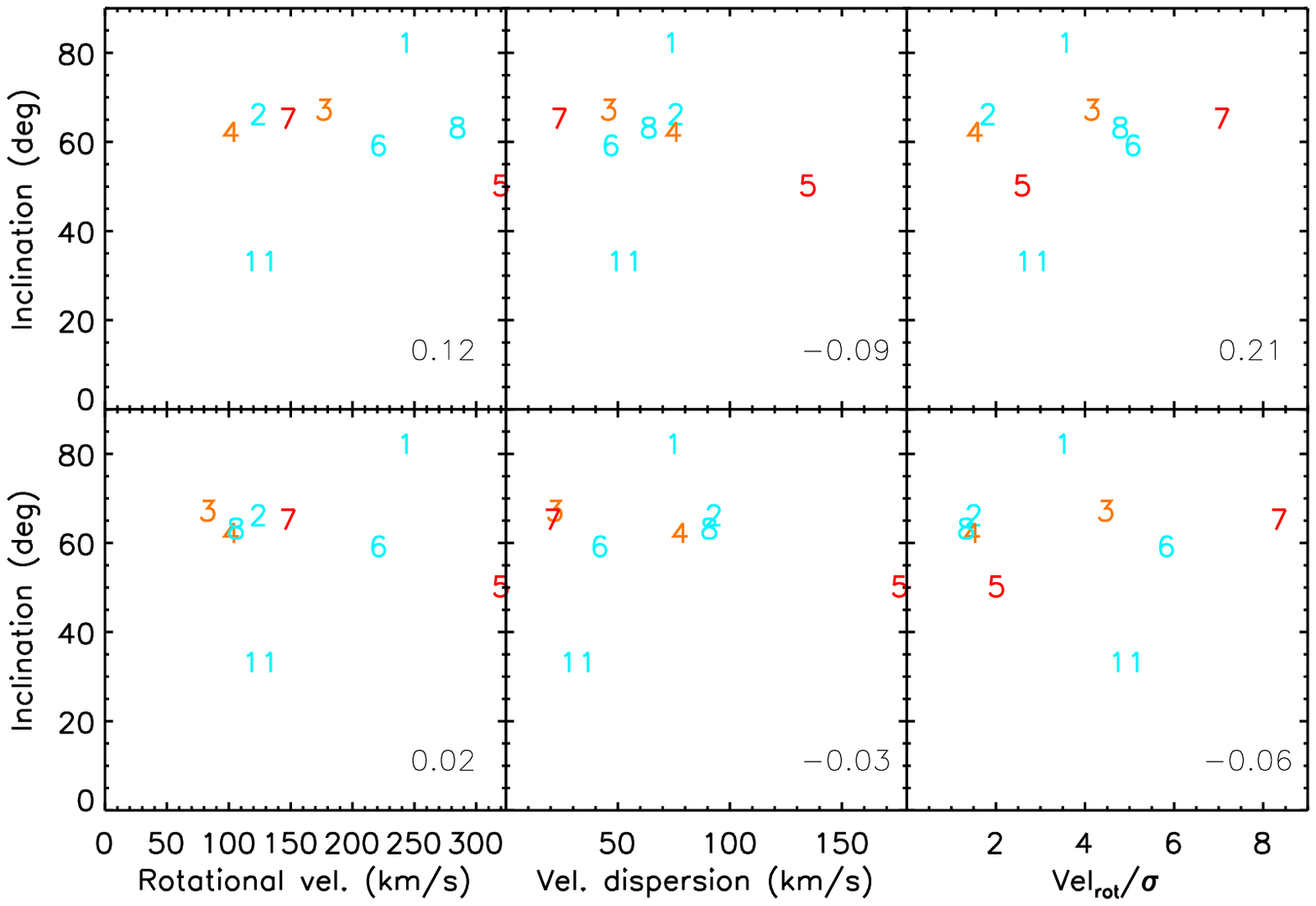}

\caption
{Inclination vs rotational velocity, $1/$error$^{2}$ velocity dispersion and $V_{max}/\sigma$ 
for each galaxy (top row) and for each galaxy within one effective radius (bottom row). Galaxies are represented by their respective numbers (for more information see Table \ref{tab:velocities}), having the
 following color coding: blue for disk-like galaxies, red for perturbed rotators and orange for interacting galaxies. Pearson correlation coefficients for each galaxy appear in the bottom right corners. In
 light of these plots, the fact that we do not find any correlation indicates there is likely no strong bias on obtaining the kinematic data for our sample based on inclination effects.}

\label{fig:inclination_correlation}

\end{figure*}

On the other hand, POWIR4 and 10 are classified as interacting objects. In fact, in both cases the \halpha emission originates (although there is a very weak detection in POWIR10 for the main object) from
 spaxels that do not belong to the target galaxy. Hence we discern two separate galaxies interacting within our SINFONI data. The photometric data has been derived for the main objects, which we identify as
 the massive objects in the \ks band imaging, while the \halpha detection comes from the secondary galaxies. These non massive galaxies are therefore excluded from our discussion and plots, but we derived
 kinematics for them to understand which physical processes are taking place in the merger. Little can be said about the two massive and main objects: POWIR4 is completely devoid of \halpha emission, while
 POWIR10 looks like a point source with a strong [NII] line in its center, that suggests it hosts an AGN. POWIR3 is an object which we include in this category as well, as its \ks and continuum images do
 not match with the \halpha emission, whose map is quite irregular. 

The rest of the objects are more difficult to catalog visually. We must bear in mind that 2 out of the 3 other objects were observed half of the nominal integration time. Either POWIR5, 7 and 11 have
 relaxed morphologies in the \ks and continuum bands while \halpha shows, as expected, a more complicated pattern. POWIR11 is different, despite the observational issues. It has an easily distinguishable
 and large rotational field, which fits better the disk object classification. The other two galaxies are catalogued as perturbed rotators. One of our major conclusions is that these massive systems have a
 wide range of properties when examined kinematically, and that many of them display significant rotational support.

\begin{figure*}
\includegraphics[angle=0,width=1.\linewidth]{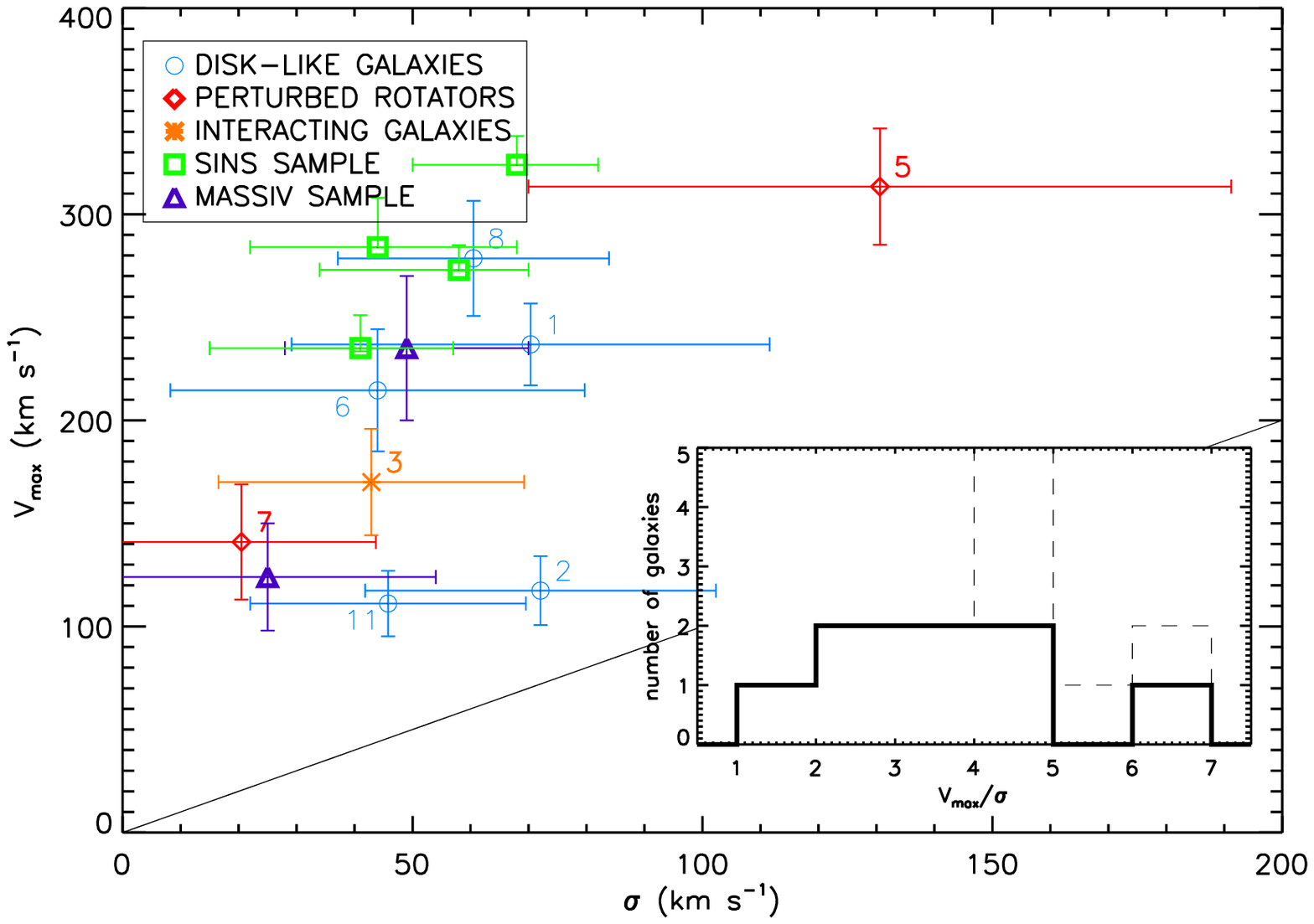}

\caption{Maximum rotational velocity inferred from our modelling versus the $1/$error$^{2}$ velocity dispersion after correcting it for beam smearing. The numbers and their symbols depict each one of the massive galaxies from
 our sample, whereas the violet triangles come from the MASSIV sample (Epinat et al. 2012) and the green squares from the SINS sample (F\"{o}rster-Schreiber et al. 2009). 
 MASSIV and SINS subsamples in our figures have been selected by means of their high stellar mass (M$_{stellar}\geq10^{11}$ h$_{70}^{-2}$ M$_{\odot}$)
 and similar redshifts (z $\sim$1-2) than our sample. Note that SINS objects error bars are asymmetric.
We also attach the histogram of the $V_{max}/\sigma$ of our massive galaxies with and without the addition of the massive galaxies in other samples (dashed or solid
 histogram respectively). For all these massive galaxies we find that $V_{max}/\sigma > 1$, as they lay above the 1:1 solid line, with most of them showing ratios 3$-$5 which corroborates their
 gravitational support. The fact that SINS and MASSIV objects lay in the upper part of this plot is further evidence that these systems are more disk-like.}
\label{fig:vmax_vs_sigma}

\end{figure*}

\subsection{Rotation vs velocity dispersion dominance}
\label{subsec:rot_vs_sigma}

In low redshift studies, the $V_{max}/\sigma$ vs. $\varepsilon$ diagram (also called the anisotropy plot; Figure \ref{fig:anisotropy_plot}) is a classical tool to measure the kinematics of early type
 galaxies (Illingworth 1977, Bender et al. 1994, Cappellari et al. 2007, Emsellem et al. 2011). We created this plot with our sample's data as an exercise, as massive galaxies at low-z are predominantly
 early type systems and therefore this is a good test to shed light into the nature of our sample. However, we remind the reader that we are dealing with ionized gas emission instead of stellar kinematics.

The plotted parameters used in this relation usually are measured at one effective radius distance from the galaxy center. To palliate our uncertainty on this, we computed effective radii in our sample
 using the relation published in Buitrago et al. (2008) for massive disk-like galaxies (to be consistent with our modelling), extrapolated to each galaxy's redshift. Then we computed $V_{max}/\sigma$ in the
 closest aperture to the calculated effective radius, as to measure velocity dispersions we need an integer number of spaxels around our kinematic center. All the information used is tabulated in Table 
\ref{tab:radii}. We add low redshift galaxies from the ATLAS$^{3D}$ Survey (Emsellem et al. 2011). Note however that their kinematics are obtained for the stellar component and that not all their masses
 fulfill our definition of a massive galaxy (M$_{*} > 10^{11}$ M$_{\odot}$). Nevertheless the comparison makes sense as their sample is composed of some of the most massive galaxies in the nearby Universe.
 Although uncertainties are large (also for the ellipticity, due to the coarse resolution of our images), we find that all of the massive galaxies at high-z from our sample lay above the line defined by
$$\left( \frac{V}{\sigma} \right) \approx 0.890 \sqrt{\frac{\varepsilon}{1-\varepsilon}}$$
\noindent which is the minimal rotational approximation to the isotropy line optimized for integral-field kinematic observations (Binney et al. 2005, Cappellari et al. 2007). This reveals the high level of
 rotational support for these massive galaxies at z$\sim$1.5, especially when comparing with slow rotators that are the most massive galaxies nearby. It is interesting as well that both POWIR2 and POWIR8
 are close to the isotropy line. Both galaxies show a clear disk with a velocity dispersion enhancement in the center, which is potentially related to a forming bulge component. Possibly these galaxies are
 beginning to transform into an early-type classification due to secular and passive evolution. However due to their total $V_{max}/\sigma$ (those not derived within one effective radius) and rotational
 velocity fields we acknowledge them as more similar to disk dominated galaxies.

\begin{figure*}
\includegraphics[angle=0,width=1.\linewidth]{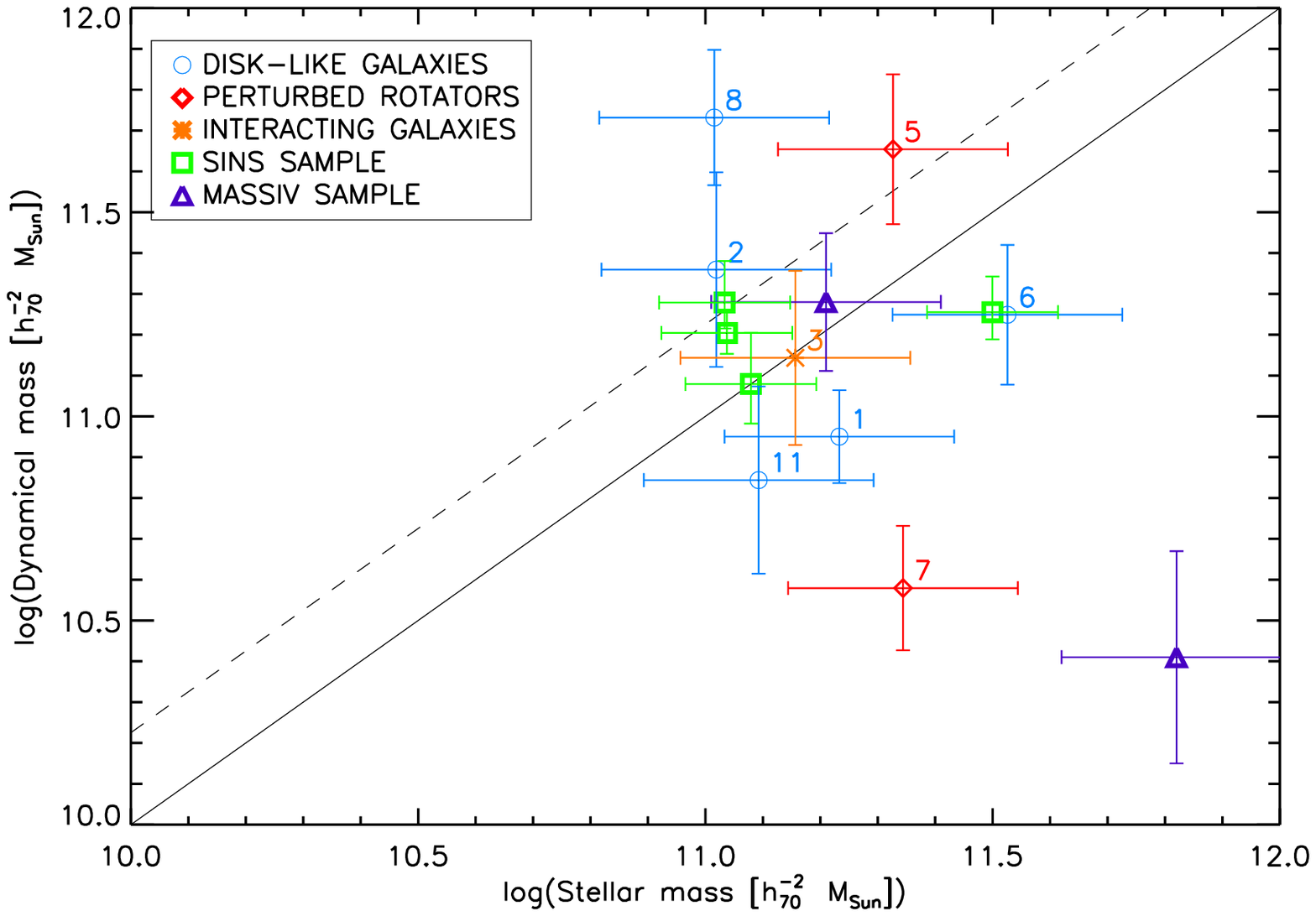}

\caption{Comparison between the inferred dynamical mass and the photometric stellar mass of our galaxies. The numbers and their symbols depict each one of the massive galaxies from our sample, whereas the violet triangles come
 from massive galaxies in the MASSIV sample (Epinat et al. 2012) and the green squares from the SINS sample (F\"{o}rster-Schreiber et al. 2009). The dynamical mass is obtained using the enclosed mass estimation due to the
 rotational velocity and adding the contribution of the velocity dispersion using the asymmetric drift correction (Meurer et al. 1996). However, for the four objects corresponding to the SINS survey,
 dynamical masses are inferred assuming the isotropic virial estimation (see section 9.6 in F\"{o}rster-Schreiber et al. 2009). The solid line is the 1:1 relationship, while the dashed line is the local
 relationship for local SDSS galaxies in Van de Sande et al. (2011). Assuming good (albeit with 0.2-0.3 dex errors) stellar mass calculations, the difference in dynamical mass may relate with the fact that
 what we measure is the inner ionized gas kinematics in our galaxies and not the overall baryonic matter contribution.}
\label{fig:masses}

\end{figure*}

In order to assess whether these findings are biased, we constrain the impact of the galaxy inclination in our sample. Figure \ref{fig:inclination_correlation} shows how our kinematic velocities (namely
 maximum rotational velocity, $1/$error$^{2}$ velocity dispersion, their respective ratio, and these values within one effective radius) are correlated according to their inclinations. Pearson correlation
 coefficients are supplied in each subplot. The small values of the Pearson coefficients confirm they do not depend on the galaxy inclination, and thus excluding this factor as a dominant source of error.

Comparing the rotational velocity versus the velocity dispersion within our systems is a good way to gauge the dynamical status of massive galaxies at high redshift, where the information is not so detailed
 as in the local Universe. This comparison is shown in Figure \ref{fig:vmax_vs_sigma} where we
 use the maximum rotational velocity from our models, and the  $1/$error$^{2}$ velocity dispersion for our sample. 
We supplemented this information with the other massive (M$_{stellar}\geq10^{11} $h$_{70}^{-2}$ M$_{\odot}$) galaxies at z $\sim$ 1-2 with SINFONI data and analysis found in the literature, namely 
the galaxies VVDS140258511 and VVDS220584167 from the MASSIV survey (Contini et al. 2012, Epinat et al. 2012) and the galaxies Q2343-BX610, D3a-6004,D3a-6397 and D3a-15504 from the SINS survey 
(F\"orster-Schreiber et al. 2009). We utilize these two samples throughout the paper and, as such, it is important to state here their selection criteria to understand any possible bias introduced by their use. 
MASSIV is a sample of star-forming galaxies at 0.9 $<$ z $<$ 2.2 based on their [OII]$\lambda$3727 or ultraviolet flux, while SINS is a compendium of previously identified (various types of color-selected objects, Lyman Break galaxies and line emitters) 
star-forming galaxies at z $\sim$ 1-4
with a general emission line flux of $\geq$ 5 $\times$ 10$^{-17}$ erg s$^{-1}$ cm$^{-2}$.

\begin{figure*}
\includegraphics[angle=0,width=1.\linewidth]{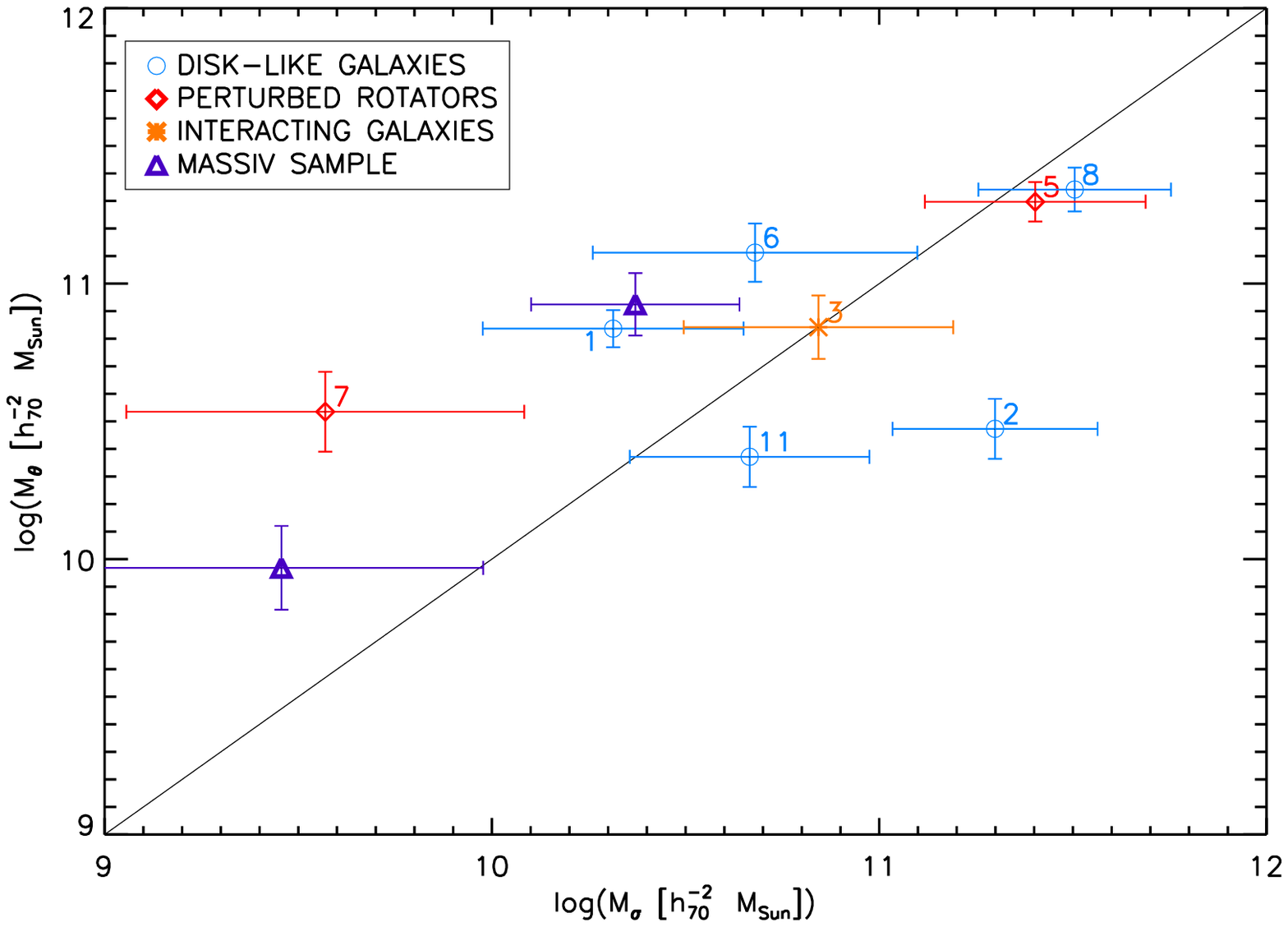}

\caption{Relative importance of the two terms which contribute to the dynamical mass (see Section \ref{subsec:dyn_masses}). The number of each galaxy is plotted, to better understand the properties of every galaxy. The solid line shows the 1:1
 relation for convenience. Note that the MASSIV massive galaxies (violet triangles, Epinat et al. 2012, correcting their masses accordingly to our Chabrier IMF) are overplotted, but not the SINS objects 
(F\"{o}rster-Schreiber et al. 2009), as their dynamical mass were calculated assuming the isotropic virial estimation (see section 9.6 in F\"{o}rster-Schreiber et al. 2009). The various kinds of galaxies
 within the present work are spread over this plot, with no obvious split in their observational properties.}
\label{fig:enclosed_mass_vs_assym_drift}

\end{figure*}

As can be seen, all the galaxies in these samples exhibit $V_{max}/\sigma>1$, in most cases with values larger than 3. We construct as well the histogram of the data shown in Figure \ref{fig:vmax_vs_sigma}.
 The dashed part corresponds to the galaxies that are not part of our sample. Although the number statistics are poor, all the massive galaxies plotted show rotational velocities exceeding their computed
 central velocity dispersions, in most cases by a large factor. Interestingly, the objects from the SINS and MASSIV surveys have $V_{max}/\sigma$ ratios which are on average larger than our values. One
 possible explanation is that, as these objects are solely selected by their star-formation, they are potentially even more rotationally dominated than our sample. This is another piece of evidence that several
 massive galaxies in our sample have settled down by $z\sim1.4$, and are developing a possible bulge component, as suggested by the anisotropy plot.

\subsection{Dynamical masses}
\label{subsec:dyn_masses}
Our integral field spectroscopy results may also be used to explore the dark matter content in our sample. To achieve this aim we computed dynamical masses combining the information coming from the
 rotational velocity and the velocity dispersion maps using the formula (from Epinat et al. 2009)
$$M_{dyn}=M_{\theta}+M_{\sigma}=\frac{V_{max}^{2}R_{last}}{G}+\frac{\sigma^{2}R_{last}^3}{Gh^{2}}$$
where $M_{\theta}$ describes the mass enclosed up to a radius $R_{last}$ and $M_{\sigma}$ is called the asymmetric drift correction (Meurer et al. 1996) by which we take into account the 1/error$^2$
 velocity dispersion ($\sigma$), i.e., the random motions within the galaxy. All of these terms are described in Table \ref{tab:masses} except $h$, which is the gas surface density disk scale length
 described by a Gaussian function, which can be expressed analytically as $h=(2\rm ln2)^{-1/2}r_{e}$. We assume that both $M_{\theta}$ and $r_{e}$ are for disk-like systems, as explained in Section 
\ref{subsec:rot_vs_sigma}. The outcomes of this calculation are plotted in Figure \ref{fig:masses}. Two lines are drawn on this figure: the solid one is the 1:1 reference, while the dashed line is the local
 relationship ($M_{dyn}/M_{*}=1.68$, the average ratio for SDSS galaxies) from Van de Sande et al. (2011). 

\begin{figure*}
\includegraphics[angle=0,width=1.0\linewidth]{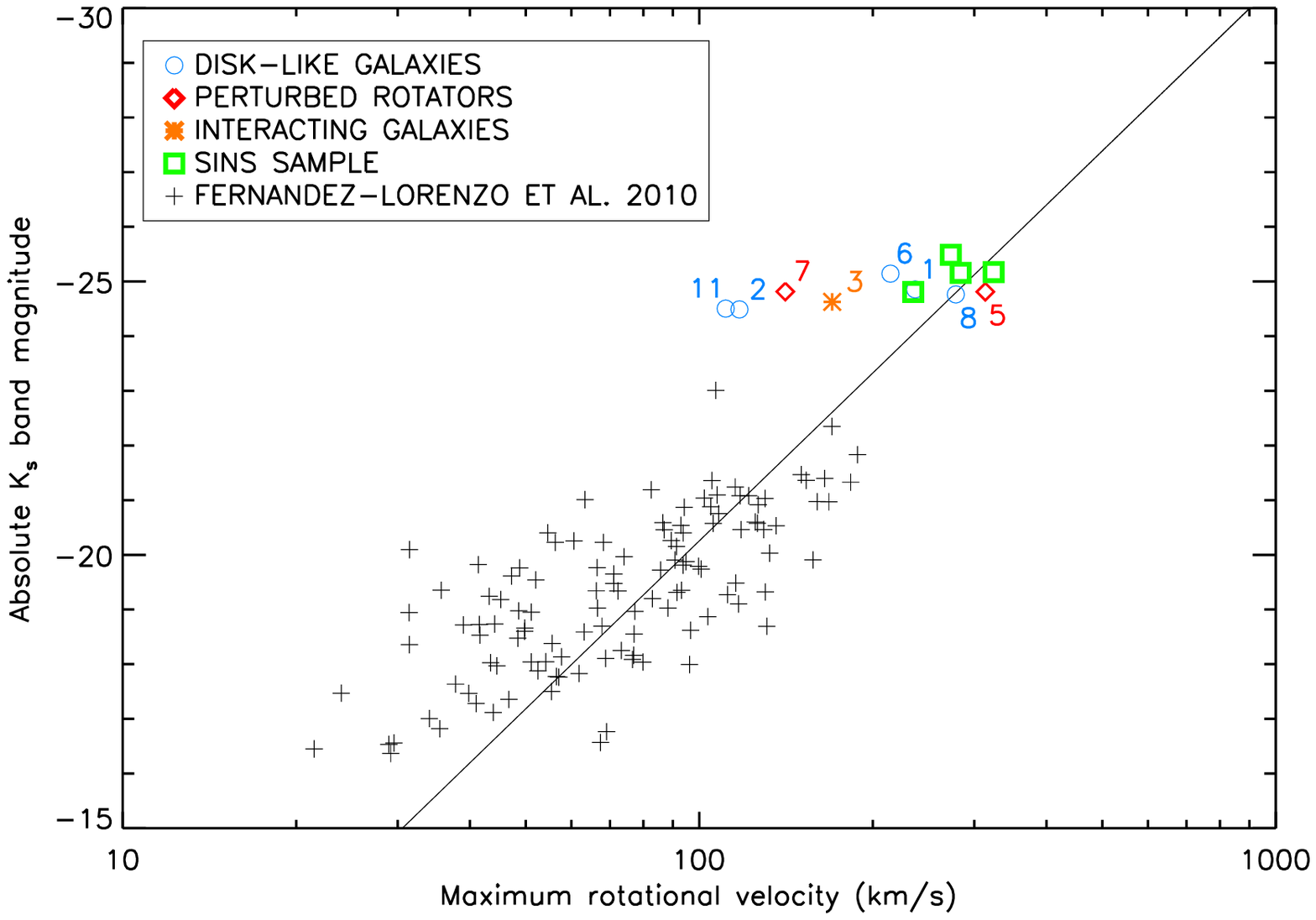}

\caption{Tully-Fisher relation with \ks absolute luminosity for our massive galaxies at high redshift. The number of each galaxy is plotted, to better understand the properties of each galaxy throughout
 the paper. Symbols have the same meaning as in previous plots, with green squares representing the massive galaxies in the SINS sample (F\"{o}rster-Schreiber et al. 2009). The solid line comes from 
Fern\'{a}ndez-Lorenzo et al. (2010), and it is the local ($z\sim0.2-0.3$) Tully-Fisher relation derived for POWIR \ks band galaxies (which the small crosses). This relationship was inferred by weighting
 the importance of every point by its errors. Galaxies displaying the lowest rotational velocity values are the ones which depart more from the local relation.}
\label{fig:abs_k_vs_v_max}

\end{figure*}

In principle, one would expect all galaxies to populate the region above the solid 1:1 line, as their dynamical mass would have to account for the baryonic mass plus the dark matter component. This is not
 the case for all the objects in our sample. There are a number of reasons which may explain this disagreement. First, we must not forget that this dynamic mass originates from the ionised gas dynamics
 which may depart from the values obtained from the stellar measurements. Secondly, our calculations account for the mass within $R_{last}$, i.e., the maximum \halpha radius, which is smaller than the
 apertures where the stellar masses have been measured, and also smaller than the typical radius used in other works such as F\"{o}rster-Schreiber et al. (2009) to obtain this parameter. Our dynamical
 masses would be larger if we correct them for these effects. Furthermore, Mart\'inez-Manso et al. (2011) claim a possible overestimation on the stellar masses we are utilising from Bundy et al. (2006) and
 Conselice et al. (2007). If confirmed, our dynamical masses would be in better agreement with the new stellar masses. Peralta de Arriba et al. (2013) conducted an investigation on the dynamical masses of 
 DEEP2 massive spheroids, which is relevant for the cases in which our disk assumption is weaker. There might also be inclination or beam smearing correction problems. 

To try to understand better the origin of our dynamical masses, we performed a plot of the relative contributions of each term in the formula used in Figure \ref{fig:enclosed_mass_vs_assym_drift}. 
Apart from the low SNR observations of POWIR7, the contribution for the velocity dispersion term is quite important for all our sample, making its addition mandatory to retrieve correct dynamical mass estimates of high-z galaxies.

\begin{figure*}
\includegraphics[angle=0,width=0.7\linewidth]{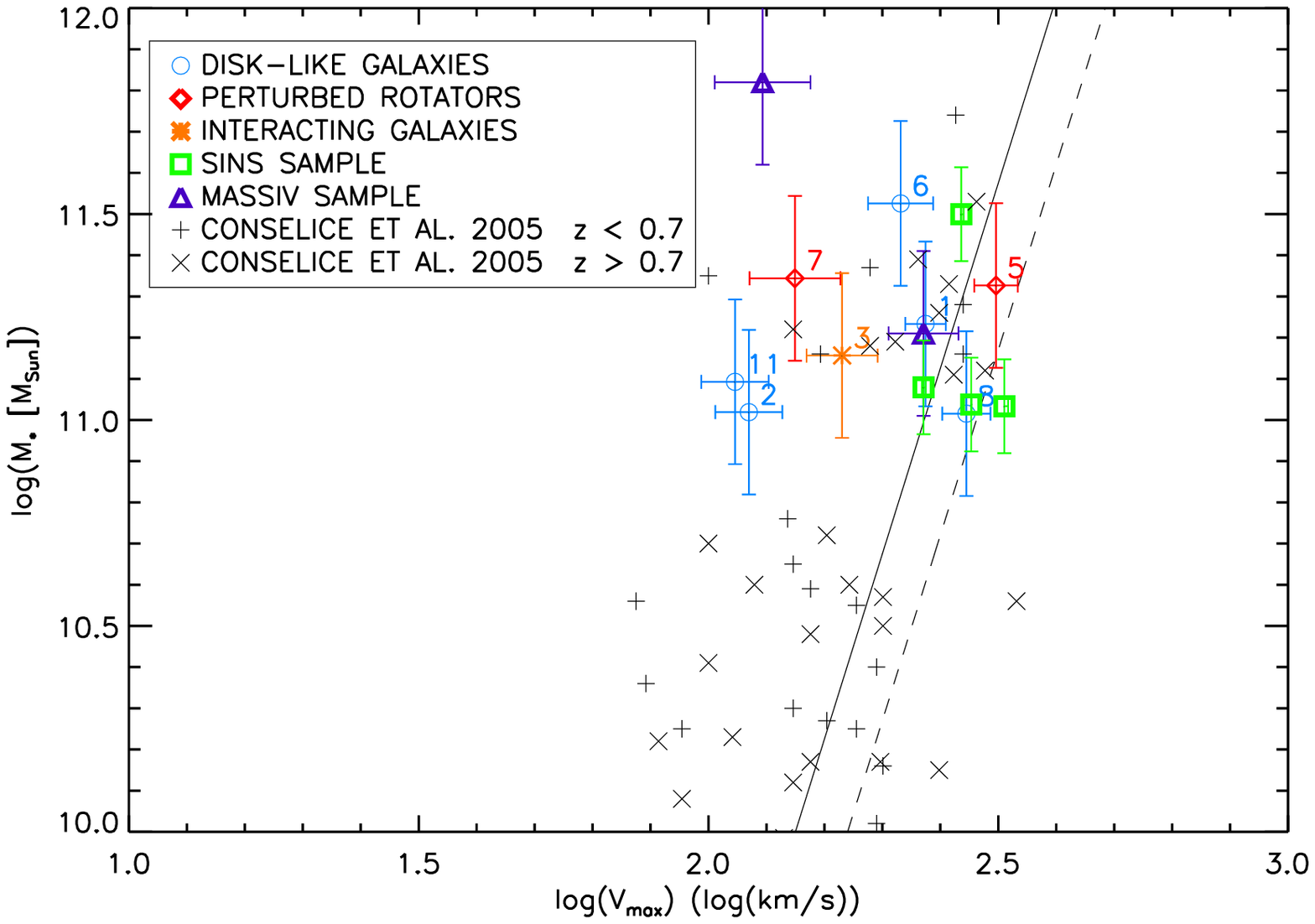}

\caption{Baryonic or stellar mass Tully-Fisher relation for our massive galaxies at high redshift. The number of each galaxy is plotted, to better understand the properties of each galaxy throughout the
 paper. To increase our number statistics for individual massive galaxies at high-z, we show massive galaxies from the MASSIV (violet triangles, Epinat et al. 2012) and SINS (green squares, F\"{o}rster-Schreiber et
 al 2009) samples as in the rest of the plots. The solid line is the local relationship from Bell \& De Jong (2001) and the dashed line is the $z=2.2$ Tully-Fisher relationship derived in Cresci et al.
 (2009) for the SINS galaxies. We also add it with the disk galaxies from Conselice et al. (2005), separating their sample between $z\leq0.7$ and $z > 0.7$ to better comprehend any possible redshift
 evolution.}
\label{fig:mass_vs_v_max}

\end{figure*}

\begin{figure*}
\includegraphics[angle=0,width=0.7\linewidth]{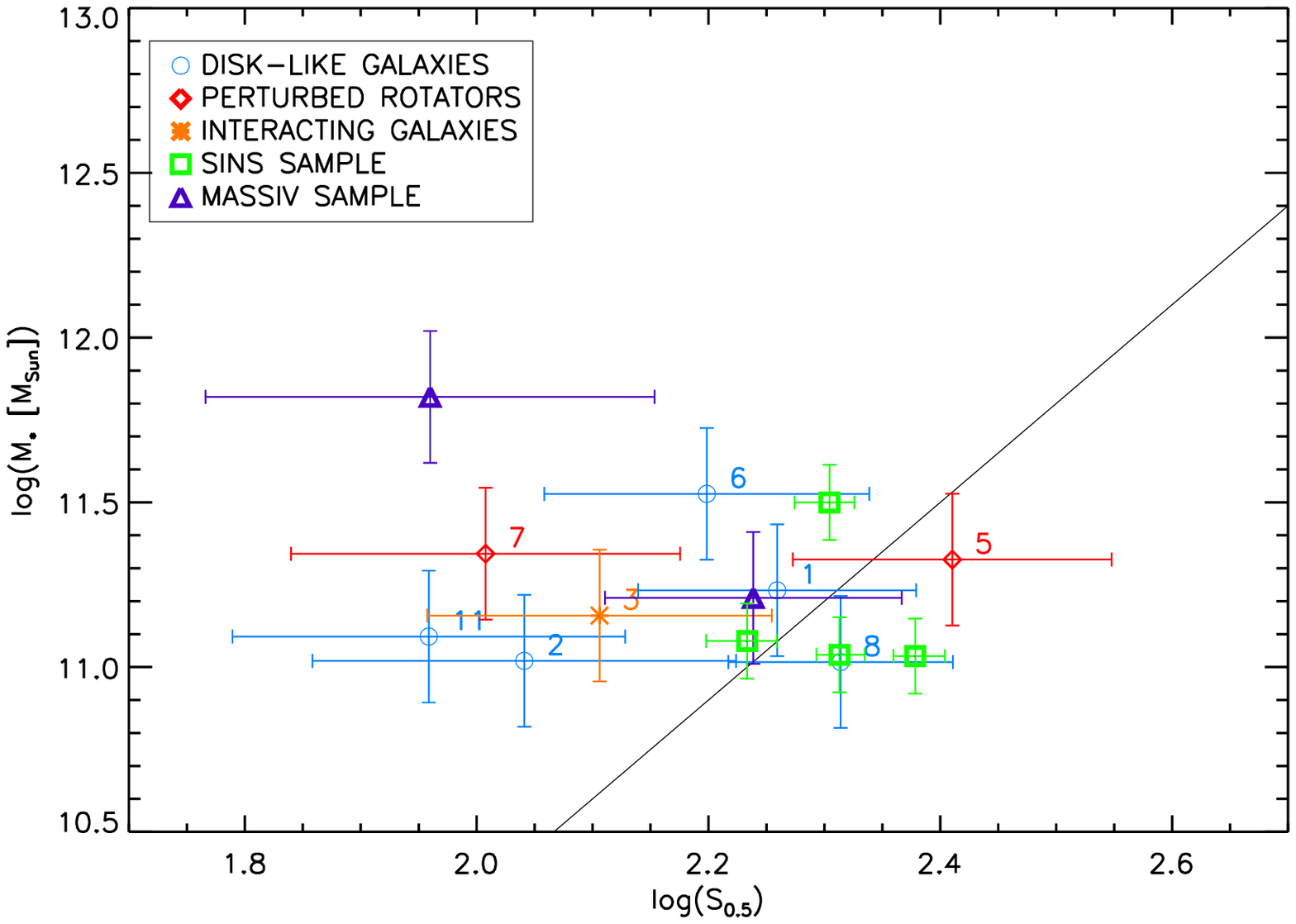}

\caption{Kassin et al. (2007) Tully-Fisher relation. On that work, the authors developed the $S_{0.5}$ parameter, which is $S_{0.5}=\sqrt{(0.5*(v_{max}^{2}))+\sigma^2}$. They argued this accounts for the
 non ordered motions of the gas and also the scatter of its Tully-Fisher relation is smaller. The solid line represents the relation they found in their closest redshift bin to our data 
($0.925 < z < 1.2$). Numbers depict each one of the massive galaxies from our sample, whereas the violet triangles come from the MASSIV sample (Epinat et al. 2012) and the green squares from the SINS
 sample (F\"{o}rster-Schreiber et al. 2009). The scatter in the massive galaxies is large, showing that the objects further away from the Kassin relationship cannot be solely described as disk-like
 galaxies.}
\label{fig:like_kassin_2007}

\end{figure*}

\subsection{Tully-Fisher relation}
\label{subsec:tully-fisher}

The Tully-Fisher Relation (TFR) links the maximum rotational velocity of spiral galaxies with their luminosity or stellar mass (Tully \& Fisher 1977; Fern\'{a}ndez-Lorenzo et al. 2009 for a comprehensive
 update). It has also been extended to S0 galaxies (e.g. Bedregal et al. 2006) and early-type galaxies in general (e.g. Davis et al. 2011). Modern investigations focus on finding and understanding any
 evolution in its slope, intercept, or both. It is a powerful scaling relation which accounts for how the stellar mass and the dark matter content are related (e.g. Conselice et al. 2005). 

Several attempts to measure the TFR with SINFONI integral-field \halpha spectroscopy have been performed. 
Van Starkenburg et al. (2008) analysed a disk galaxy at z = 2.03 which followed the local K-band (Verheijen 2001) relation. Cresci et al. (2009) found at $z\sim2.2$ a slope consistent with the Bell \& de Jong (2001)
 $z\sim0$ relationship. At similar redshifts than our sample, i.e. 1 $\leq$ z $\leq$ 1.5, both Lemoine-Busserolle \& Lamareille (2010) and Vergani et al. (2012) were compatible with Bell \& de Jong relation, but with a large scatter owing to having dispersion dominated/low rotational velocity galaxies.

We attempt to shed some light into the TFR for our sample. The number of objects is not high, and their morphological nature is not perfectly constrained, but in spite of these limitations this is one of
 the few cases where this relationship could be obtained for IFU observations of massive galaxies at high redshift. Moreover, it helps understanding how the mass and the rest of kinematic properties of our
 sample are connected. 

We show the \ks band TFR for our sample of galaxies in  Fig. \ref{fig:abs_k_vs_v_max} using the maximum rotational velocity retrieved in our modelling. As we did in the previous plots, we add SINFONI
 \halpha measurements for the massive galaxies in the SINS survey (F\"{o}rster-Schreiber et al. 2009, with a detailed TFR study in the aforementioned Cresci et al. 2009). The solid line accounts for the
 local ($z\sim0.2-0.3$) \ks band POWIR/DEEP2 relationship found in Fern\'{a}ndez-Lorenzo et al. (2010), with the crosses the objects studied in that article. Overall, our results are similar to Epinat et
 al. (2009) and Vergani et al. (2012), as the galaxies which display the lowest rotational velocities (POWIR2 and POWIR11) are the ones further from the fiducial relation. When plotting the results for
 POWIR4 and POWIR10, they are placed in a similar locus, but the reason for this would be their allegedly non-massive nature
. Whereas for POWIR2 this constitutes a further indication of its departure from a pure disk system, the interpretation is not so obvious for POWIR11, when looking also at the anisotropy plot (Fig.
 \ref{fig:anisotropy_plot}). We attribute this to the fact that this is the object with the lowest inclination in our sample, and subsequently it is more difficult to constrain this parameter which affects
 the rotational velocity determination.

The stellar mass or baryonic TFR (using stellar mass instead of luminosity) has been claimed to be a better proxy for studying galaxy mass assembly than the usual TFR. We show the baryonic TFR for our
 sample in Fig. \ref{fig:mass_vs_v_max}. The solid line is the local relationship found by Bell \& de Jong (2001), corrected to our Chabrier IMF, while the dashed line is the derived stellar mass TFR at 
$z\sim2.2$ in Cresci et al. (2009). We also add the disk galaxies from Conselice et al. (2005), separating their sample between $z\leq0.7$ and $z > 0.7$ as they did. There is a close resemblance with Fig.
 \ref{fig:abs_k_vs_v_max}. We can see that all the massive galaxies in the present study occupy similar loci, giving further evidence for our disk assumption in the galaxy modelling. However, 
 any redshift trend is unclear in the data, as it occurred in Kassin et al. (2012). In that latter study, the authors plotted a series of galaxy samples with reliable spectra up to z $\sim$ 3, being the
 last the ones pertaining to the AMAZE/LSD survey (Gnerucci et al. 2011). We must bear in mind that these objects escape from the scope of the paper, as this is a SINFONI programme which aimed at characterizing very high redshift star-forming galaxies.

In order to disentangle better the disordered motions of the gas we follow the prescriptions in Kassin et al. (2007) -- see also Cresci et al. (2009), Lemoine-Busserolle \& Lamareille (2010) and Vergani et al. (2012) --, where they used the
 parameter $S_{0.5}=\sqrt{(0.5*V_{max}^{2})+\sigma^2}$, arguing that the scatter in the TFR is tighter when taking into account the contribution in the velocity dispersion. We show this relation in Fig.
 \ref{fig:like_kassin_2007}, plotting the highest redshift ($0.925 < z < 1.2$) relation inferred in Kassin et al. (2007). There does not appear to be any conspicuous difference from the rest of TFR plots,
 suggesting again that the massive galaxies with the lowest values of maximum rotational velocity have properties that are more difficult to match with the assumption of a pure disk-like nature.

\section{Discussion \& Conclusions}
\label{sec:conclusions}

We present a SINFONI study of a sample of massive galaxies (M$_{stellar}\geq10^{11}$ h$_{70}^{-2}$ M$_{\odot}$) at $z\sim1.4$ selected by their stellar mass and emission lines (with EW[OII] $> 15$ \AA), in
 order to understand the kinematics/secular motions of this galaxy population and to constrain their rotational nature. This is a matter of debate after recent photometric studies (e.g. van der Wel et
 al. 2011, Buitrago et al. 2013 among others) found that high redshift massive galaxies are predominantly disk like.  There are also \halpha detections using slitless spectroscopy (cf. 3D-HST Survey; van
 Dokkum et al. 2011) showing intense star formation in massive high-z galaxies. Our work is an attempt to clarify the diversity of properties within these galaxies, and whether they are better described
 kinematically by a disk-like or a spheroidal-like population. 

In this study we have carefully chosen 10 massive galaxies with available deep Keck spectroscopy and \ks band imaging from the POWIR/DEEP2 survey. VLT/SINFONI H-band observations  with a good (average
 $0.56$ arcsec) seeing, enabled us to build \halpha kinematic maps. We fit rotating disk models to their velocity fields that allowed us to derive rotation velocities and correct the velocity dispersion
 maps from beam smearing. Hence we minimize potential error sources as the broadening of the spectral lines by velocity shear. 

An open question is whether we could generalize our conclusions to the whole massive galaxy population at z$\sim$1.4. To answer this we need first to ascertain whether our sample is unusual for showing 
[EW]$> 15$ \AA, and thus non negligible star formation rates, as both \halpha and [OII] come from ionized gas by star formation in HII regions. Among studies similar to ours, but using traditional slit spectroscopy, Twite et al. (2012) analyse
 \halpha star formation in M$_{stellar} >$10$^{10.5}$ M$_{\odot}$ galaxies at 0.4 $<$ z $<$ 1.4. Although their sample was not complete in stellar mass, they found a significant drop at lower redshift in
 the number of massive galaxies with detected \halpha emission. 


Although they do not sample the same redshift range as our study, Sobral et al. (2011) find that the  the fraction of massive galaxies with M$_{stellar}\geq10^{11}$ M$_{\odot}$ detected in H$\alpha$ is 
$\sim15$\% at $z = 0.84$.  These systems have equivalent widths greater than $\sim$15 \AA, which typically translates into star formation rates of 5-10 M$_{\odot} {\rm yr}^{-1}$ (P. Best private
 communication). Finally, Fumagalli et al. (2012), working with slitless spectroscopy in the 3D-HST survey, have found (based on their Figure 2) that at z$\sim$1.4 massive galaxies typically show
 equivalent widths similar to the ones in our sample. Summarizing, despite (on average) a level of star formation and thus a certain amount of \halpha emission is expected, our conclusions should be taken
 carefully as it is not perfectly clear yet whether all massive galaxies have the same characteristics of our sample at this specific redshift (z$\sim$1.4).

Half of our sample's members have been identified as rotating disks (POWIR1, 2, 6, 8 and 10), based on their ordered rotational velocity gradients. All galaxies from our sample show $V_{max}/\sigma > 1$, where this ratio is on average 3$-$4 in most cases (Figure \ref{fig:vmax_vs_sigma}). The velocity dispersions, although smaller than
 the typical values reported for the stellar component in massive galaxies at these redshifts (see for example Cenarro \& Trujillo 2009), resemble those found for ionised gas kinematics. These large 
$V_{max}/\sigma$ ratios are however at odds with local Universe counterparts, which either display $V_{max}/\sigma < 1$ (e.g. Emsellem et al. 2011) for early-type galaxies or $V_{max}/\sigma \sim 6-14$
 (e.g. Epinat et al. 2010) in case of IFU studies of \halpha emission in spirals. 

We agree with the explanation given in previous high redshift 3D spectroscopy studies (Bournaud et al. 2008, Starkenburg et al. 2008, Law et al. 2009, F\"{o}rster-Schreiber et al. 2009, Lemoine-Busserolle \& Lamareille 2010, Lemoine-Busserolle et al. 2010, Gnerucci et al. 2011, Epinat et al. 2012) such that, at high redshift, galaxy formation
 and evolution is a more turbulent process because of the larger amounts of cold gas involved, which at the same time leads to higher star formation rates than in the present day Universe. Likewise, we
 observe that major merging is indeed occurring in our sample (see POWIR4 or POWIR10 galaxies). However most of the gas should be accreted either via minor merging, whose hints are found in multiple
 galaxies of our sample, or cold gas flows along cosmic web filaments (e.g. Conselice et al. 2013).

The main difference between our sample and previous published datasets (such as the MASSIV, AMAZE/LSD or SINS surveys) lays in our high stellar mass selection (M$_{stellar}\geq10^{11} h_{70}^{-2}$ M$_{\odot}$).
 Observationally, we find that our sample consists of quite regular velocity fields showing high rotation. As stated in Epinat et al. (2012) when discussing their disk galaxies, this fact implies that the
 most massive disks seem to be stable objects even at early cosmic times. 
We have compared our massive galaxy sample with other galaxies in the literature selected by mass.  Strikingly, the conclusions remain the same. In addition, we have noticed that less-massive galaxies (Law
 et al. 2007, Wright et al. 2007, F\"{o}rster-Schreiber et al. 2009, Lemoine-Busserolle \& Lamareille 2010, Gnerucci et al. 2011, Epinat et al. 2012) contain a higher percentage of clumpy or distorted objects than our sample.

We conclude that massive galaxies may acquire more rapidly a given morphology and gravitational equilibrium than less-massive objects, accounting for a type of morphological downsizing. 
This conclusion is reinforced by Kassin et al. (2012) making use of similar DEEP2 data and obtaining that the more massive (10.4 $<$ log M / M$_{\odot}$ $<$ 10.7) disks in their sample have higher rotational velocity and less disordered motions (using their own words, they are more kinematically settled) than the less massive disks. We attribute this
 characteristic to the fact that, as we are dealing with the most massive galaxies at these redshifts, their high masses help protect them from being perturbed. Stellar mass has not only a profound impact
 in the galactic assembly and the star formation history of these galaxies, but it is also crucial for understanding their eventual morphological development, whereby they progressively obtain the observational properties of the massive galaxies we find in the nearby Universe. 

Future NIR high resolution photometry over larger samples of massive galaxies shall contribute to corroborate this scenario. These objects should be the base of surveys taking advantage of new generation multi-IFU spectrographs which will increase by a high factor the number of galaxies with available kinematic information.

\section{Acknowledgements}
\label{sec:ack}
We thank to the anonymous referee whose ideas improve the contents of the present article.
We are indebted with Nacho Trujillo for his support and ideas for this paper. Mirian Fern\'andez-Lorenzo kindly shared with us the data for the local sample in Figure \ref{fig:abs_k_vs_v_max}. We would
 like to acknowledge Boris Ha\"{u}\ss ler and Jes\'{u}s Varela for their help in better understanding the data, and also Jes\'us Falc\'on-Barroso, Nicolas Bouch\'{e}, Michele Cappellari and Ignacio
 Ferreras for their advice. We gratefully thank Thierry Contini and the University of Nottingham to facilitate BE's trip to Nottingham. We show our gratitude to the ESO SINFONI staff (Andrea Modigliani,
 Konstantin Mirny and Dieter N\"urnberger) for their assistance. FB acknowledges the support of the European Research Council and CJC both the Leverhulme Trust and the STFC. Funding for the DEEP2 Galaxy Redshift Survey has been provided by NSF grants AST-95-09298, AST-0071048, AST-0507428, and AST-0507483 as well as NASA LTSA grant NNG04GC89G.

\begin{table*}
\begin{minipage}{\textwidth}
\caption{Modelled kinematic parameters of our sample}
\label{tab:velocities}
\resizebox{\textwidth}{!}{
\begin{tabular}{c|c|c|c|c|c|c|c|c}
     Name & Velocity dispersion & Max. rotational velocity & $V_{max}/\sigma$ & Vel. disp. in $r_{e}$ & Max. rot. vel. in $r_{e}$ & $(V_{max}/\sigma)_{e}$ & Inclination        & Classification\\
          & km s$^{-1}$         & km s$^{-1}$              &                  & km s$^{-1}$ & km s$^{-1}$ &     & $^\circ$ (degrees)  & \\
     (1)  & (2)                 & (3)                      & (4)              & (5)         & (6)         & (7) & (8)                 & (9)\\
    \hline
POWIR1 & $  70 \pm 41 $ & $  236 \pm 19 $ & $ 3.37 \pm 0.29 $ & $  71  \pm 40  $ & $ 236 \pm 15  $ & $  3.32 \pm 0.21 $ & $ 80 $ & D\\ 
POWIR2 & $  72 \pm 30 $ & $  117 \pm 16 $ & $ 1.63 \pm 0.24 $ & $  89  \pm  7  $ & $ 117 \pm  9  $ & $  1.31 \pm 0.09 $ & $ 64 $ & D\\
POWIR3 & $  42 \pm 26 $ & $  170 \pm 25 $ & $ 3.96 \pm 0.62 $ & $  18  \pm 29  $ & $  76 \pm 19  $ & $  4.27 \pm 1.14 $ & $ 65 $ & I\\
POWIR4 & $  71 \pm 30 $ & $   95 \pm 35 $ & $ 1.34 \pm 0.50 $ & $  74  \pm 22  $ & $  95 \pm 13  $ & $  1.28 \pm 0.17 $ & $ 60 $ & I \\  
POWIR5 & $ 131 \pm 60 $ & $  313 \pm 28 $ & $ 2.40 \pm 0.22 $ & $  172 \pm 66  $ & $ 313 \pm 37  $ & $  1.82 \pm 0.21 $ & $ 48 $ & P \\
POWIR6 & $  43 \pm 35 $ & $  214 \pm 29 $ & $ 4.88 \pm 0.69 $ & $  38  \pm 32  $ & $ 214 \pm 28  $ & $  5.63 \pm 0.75 $ & $ 57 $ & D \\
POWIR7 & $  20 \pm 23 $ & $  141 \pm 27 $ & $ 6.88 \pm 1.42 $ & $  17  \pm 22  $ & $ 141 \pm 26  $ & $  8.16 \pm 1.58 $ & $ 63 $ & P \\
POWIR8 & $  60 \pm 23 $ & $  278 \pm 27 $ & $ 4.60 \pm 0.47 $ & $  87  \pm 18  $ & $  99 \pm 21  $ & $  1.14 \pm 0.24 $ & $ 61 $ & D \\
POWIR10 & $ 59 \pm 26 $ & $  113 \pm 18 $ & $ 1.92 \pm 0.32 $ & $  53  \pm 22  $ & $ 114 \pm 10  $ & $  2.13 \pm 0.19 $ & $ 60 $ & I \\    
POWIR11 & $ 45 \pm 23 $ & $  111 \pm 15 $ & $ 2.43 \pm 0.36 $ & $  24  \pm 22  $ & $ 111 \pm 10  $ & $  4.53 \pm 0.46 $ & $ 31 $ & D \\
\hline
\end{tabular}
}\\
Notes. (1) Name of the galaxy (2) `1/error$^2$'-weighted average velocity dispersion from the modelled velocity dispersion after removing the beam smearing (3) Maximum rotational velocity from our
 rotational velocity modelling (4) Ratio between the maximum rotational velocity and velocity dispersion (5) `1/error$^2$'-weighted average velocity dispersion from the modelled velocity dispersion within
 one effective radius after removing the beam smearing (6) Maximum rotational velocity at one effective radius from our rotational velocity modelling (7) Ratio between the maximum rotational velocity and
 velocity dispersion at one effective radius (8) Inclination as measured from GALFIT analysis. Note that for POWIR4 and POWIR10 values are fixed to 60$^\circ$, as for these two cases the observed objects
 are thought not to be the targeted massive galaxies (9) Final kinematic classification for our massive galaxies: D for rotating Disks, I for Interacting galaxies and P for Perturbed rotators.
\end{minipage}
\end{table*}

\begin{table*}
\begin{minipage}{\textwidth}
\caption{Radii used in our calculations}
\label{tab:radii}
\resizebox{\textwidth}{!}{
\begin{tabular}{c|c|c|c|c|c|c|c}
    Name & Model radius            & Model radius           & \halpha maximum radius & \halpha maximum radius & Effective radius & Pixels taken & Radius taken \\
         & arcsec                  & kpc                    & arcsec                 & kpc                    & kpc              &              & kpc     \\
    (1)  & (2)                     & (3)                    & (4)                    & (5)                    & (6)              & (7)          & (8)     \\
    \hline
POWIR1  & $ 0.12 $ & $ 1.05  $ & $ 0.62 $ & $  5.26  $ & $    3.36  $ & $  3.5   $ & $ 3.68 $ \\
POWIR2  & $ 0.12 $ & $ 1.05  $ & $ 1.10 $ & $  9.27  $ & $    2.59  $ & $  2.5   $ & $ 2.63 $ \\
POWIR3  & $ 0.70 $ & $ 5.90  $ & $ 1.22 $ & $ 10.30  $ & $    3.06  $ & $  2.5   $ & $ 2.63 $ \\
POWIR4  & $ 0.12 $ & $ 1.05  $ & $ 1.05 $ & $  8.83  $ & $    4.18  $ & $  3.5   $ & $ 3.69 $ \\
POWIR5  & $ 0.43 $ & $ 3.64  $ & $ 1.03 $ & $  8.68  $ & $    3.77  $ & $  3.5   $ & $ 3.68 $ \\
POWIR6  & $ 0.16 $ & $ 1.35  $ & $ 1.44 $ & $ 12.10  $ & $    4.81  $ & $  4.5   $ & $ 4.74 $ \\
POWIR7  & $ 0.12 $ & $ 1.05  $ & $ 0.88 $ & $  7.41  $ & $    3.85  $ & $  3.5   $ & $ 3.69 $ \\
POWIR8  & $ 0.88 $ & $ 7.43  $ & $ 1.44 $ & $ 12.17  $ & $    2.58  $ & $  2.5   $ & $ 2.63 $ \\
POWIR10 & $ 0.12 $ & $ 1.05  $ & $ 1.06 $ & $  8.90  $ & $    2.87  $ & $  2.5   $ & $ 2.63 $ \\
POWIR11 & $ 0.12 $ & $ 1.05  $ & $ 0.97 $ & $  8.19  $ & $    2.83  $ & $  2.5   $ & $ 2.63 $ \\
\hline
\end{tabular}
}\\
Notes. (1) Name of the galaxy (2) Radius of our kinematic model in kpc (3) Radius of our kinematic model in arcsec (4) \halpha maximum extent in arcsec (5) \halpha maximum extent in kpc (6) Effective
 radius as calculated by the disk-like relation for massive galaxies in Table 2 of Buitrago et al. (2008) (7) Pixels taken as effective radius according to previous column (note that the 0.5 is added as we start from the kinematic center). (8) Equivalent in kpc of the previous column.
\end{minipage}
\end{table*}

\begin{table*}
\begin{minipage}{\textwidth}
\caption{Masses inferred for our sample \& N2 calibrator}
\label{tab:masses}
\resizebox{\textwidth}{!}{
\begin{tabular}{c|c|c|c|c|c|c}
   Name & Stellar mass            & Dynamical mass           & Mass enclosed in \halpha max. radius & Asymmetric drift correction & N2 calibrator & Notes about N2 calibrator\\
        & $10^{10} h_{70}^{-2} M_{\odot}$ & $10^{10} h_{70}^{-2} M_{\odot}$  & $10^{10} h_{70}^{-2} M_{\odot}$ & $10^{10} h_{70}^{-2} M_{\odot}$  &      &       \\
    (1) & (2)                     & (3)                      & (4)                               & (5)                         & (6)           & (7)            \\
    \hline
POWIR1 & $   17.1$ & $    9.7$ &  $   6.9$  & $    2.8$ & -     & OH line over $[NII]6584$\AA wavelength  \\
POWIR2 & $   10.4$ & $   30.6$ &  $   3.0$  & $   27.6$ & -0.31 & \\
POWIR3 & $   14.3$ & $   16.6$ &  $   6.9$  & $   9.7$  & -     & OH line over $[NII]6584$\AA wavelength   \\
POWIR4 & $   25.7$ & $   10.9$ &  $   1.9$  & $   9.0$  & -0.59 &  \\
POWIR5 & $   21.2$ & $   54.9$ &  $   19.8$ & $   35.0$ & -0.22 & \\
POWIR6 & $   33.6$ & $   19.6$ &  $   13.0$ & $   6.6$  & -0.31 & OH sky line residual increases this ratio \\
POWIR7 & $   22.1$ & $    3.9$ &  $   3.4$  & $  0.5$   & -     & OH line over $[NII]6584$\AA wavelength   \\
POWIR8 & $   10.4$ & $   66.3$ &  $   22.0$ & $   44.3$ & -0.60 & \\
POWIR10 & $   12.7$ & $   16.0$ & $   2.7$  & $   13.3$ & -0.33 & The value for the massive galaxy is N2 = $-0.15$\\
POWIR11 & $   12.4$ & $    8.8$ & $   2.4$  & $   6.4$  & -0.38 & \\
\hline
\end{tabular}
}\\
Notes. (1) Name of the galaxy (2) Stellar mass from the parent POWIR/DEEP2 survey (3) Dynamical mass, as calculated in Section \ref{subsec:dyn_masses} (4) Enclosed mass term, as calculated in Section 
\ref{subsec:dyn_masses} (5) Asymmetric drift correction, as calculated in Section \ref{subsec:dyn_masses} (6) N2 calibrator as in Queyrel et al. (2009) (7) Observational notes
\end{minipage}
\end{table*}


%
%


\appendix
\section{Individual galaxy observations} 
\label{sec:galaxies}

\begin{figure*}
\includegraphics[angle=90,width=0.4\linewidth]{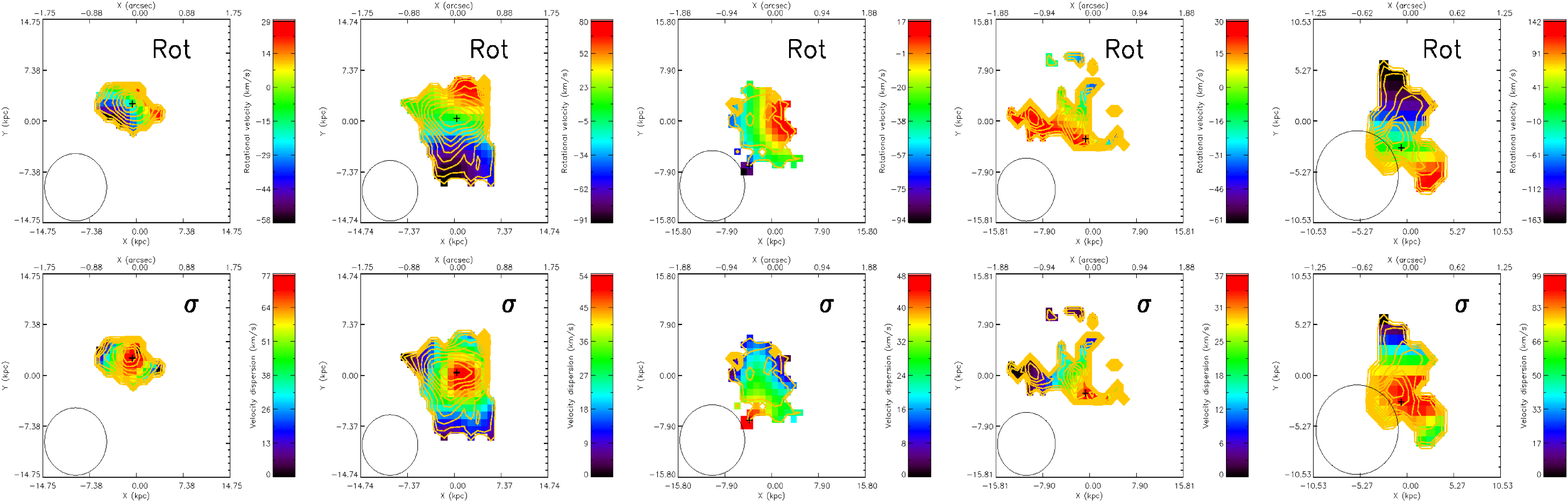}

\caption{Montage with the kinematical models. First row are the rotational velocity maps, and the second the velocity dispersion maps. From bottom to top, the columns are POWIR1, POWIR2, POWIR3, POWIR4, and POWIR5. Please note that the full descriptions of our maps are written in the Appendix A and in the caption of the first galaxy montage (Figure \ref{fig:montage_powir1}).}
\label{fig:montage_models1} 

\end{figure*}

\begin{figure*}
\includegraphics[angle=90,width=0.33\linewidth]{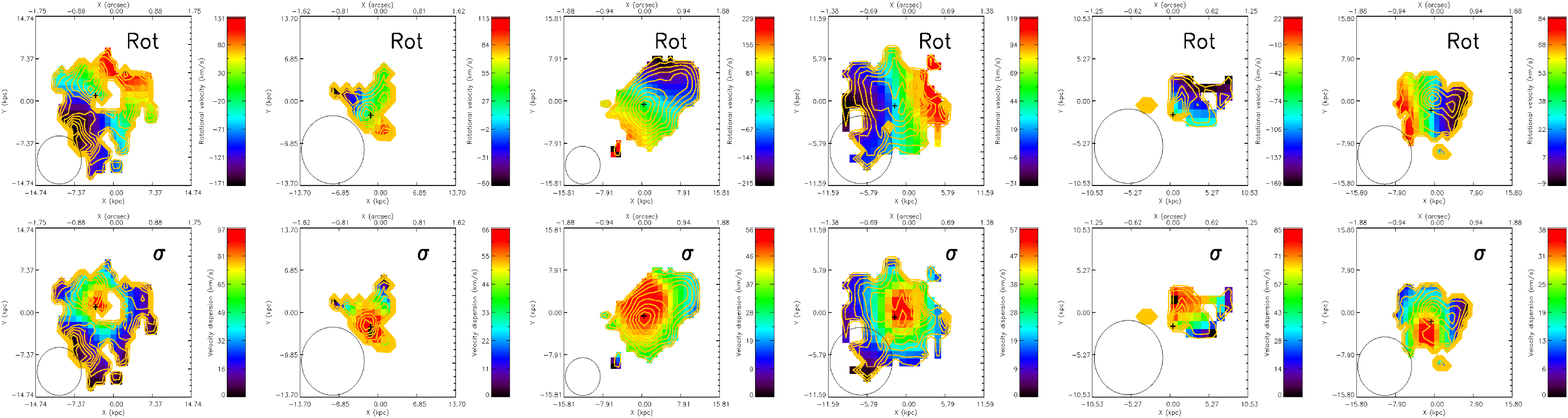}

\caption{Montage with the kinematical models for the rest of the massive galaxies not plotted in the previous Figure \ref{fig:montage_models1} . First row are the rotational velocity maps, and the second the velocity dispersion. From bottom to top, the columns are POWIR6, POWIR7, POWIR8, POWIR10 (non-massive galaxy), POWIR10 (massive galaxy) and POWIR11.}
\label{fig:montage_models2}

\end{figure*}

In this appendix we present a detailed description of each massive galaxy within our survey and their observed kinematics. Before going into this, we preface this discussion by showing the model maps in
 Figures \ref{fig:montage_models1} and \ref{fig:montage_models2}. Each galaxy explanation compares the various maps (cf. Figures \ref{fig:montage_powir1} - \ref{fig:montage_powir11}), namely the parent
 POWIR/DEEP2 \ks~band imaging, the \halpha flux images from SINFONI, H-band continuum, SNR map, radial velocity (and its residual), observed velocity dispersion (and the inferred velocity dispersion after
 removing the beam smearing). 


In our maps, north is up and east is left. \halpha contours are overplotted in all of the maps (with decrements of 10\% in flux between adjacent contours), in order to facilitate the reader in determining
 which spaxels belong to the galaxies. The axes show sizes both in kpc and arcsec. The kinematic centers used for our models are located in the spaxels with the maximum flux in the continuum maps, and are
 highlighted by a cross. Seeing values are presented by circles with their FWHM corresponding size in the left bottom corner of the SINFONI maps. In the kinematic maps, the coloured spaxels shown are all
 above a certain threshold in SNR which is written in Table \ref{tab:structural_data},  Column (S/N)$_{threshold}$. However, this threshold was not applied for the continuum maps, in order to understand the
 origin of the \halpha emission. Finally, we attempted to quantify the existence of AGN sources within our sample by means of obtaining the N2 indicator -- the logarithm of the $[NII]/$\halpha ratio; see
 Queyrel et al. (2009) for details --. For POWIR1, 3 and 7 this could not be accomplished as the [NII]$\lambda6583$\AA line is located over OH sky lines (see Table \ref{tab:masses}). We proceed to
 describe any interesting feature of our sample in the following galaxy subsections.


\subsection{POWIR1} 
\label{subsec:powir1}
\begin{figure*}
\includegraphics[angle=0,width=0.70\linewidth]{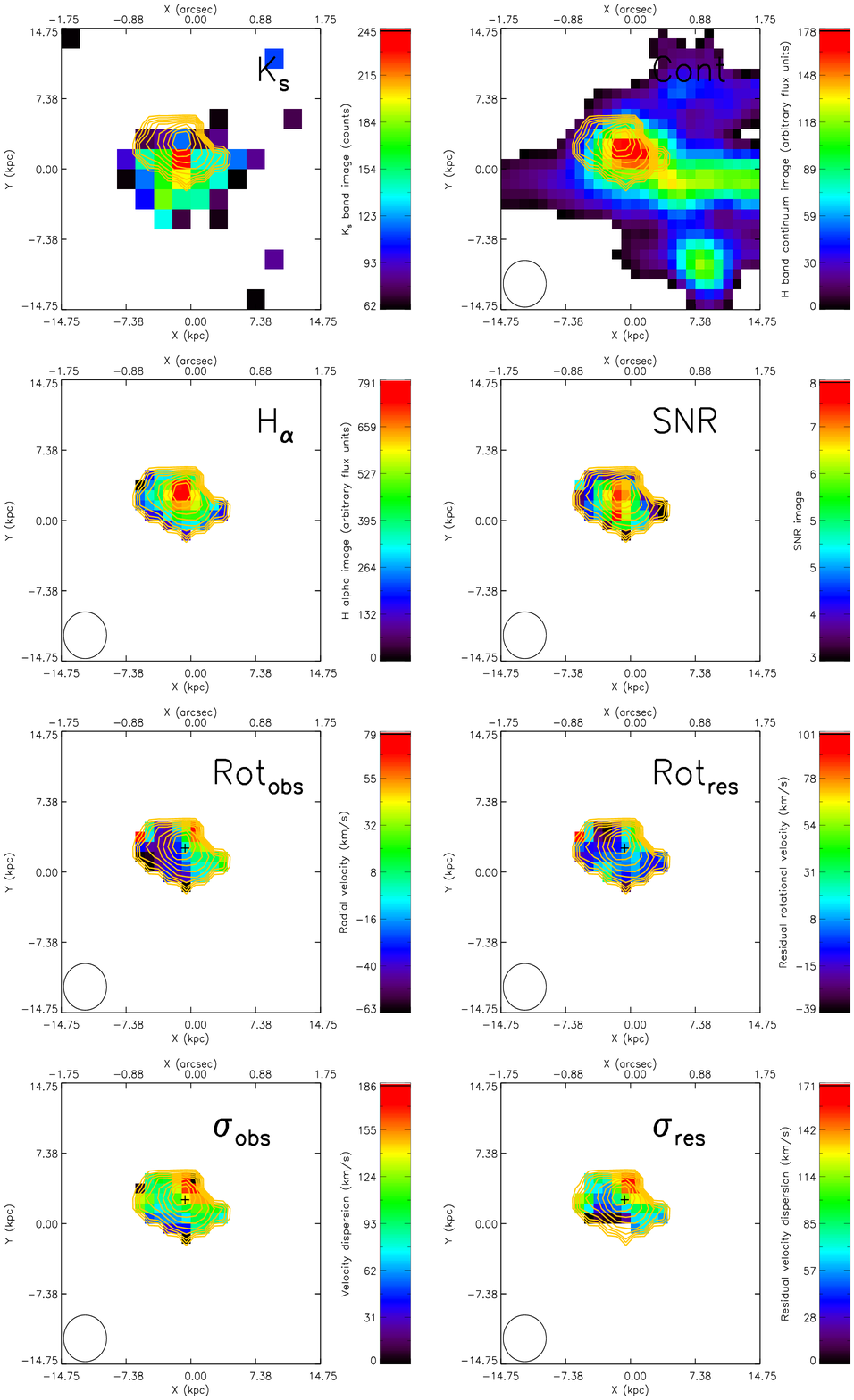}

\caption{From left to right, from top to bottom: \ks POWIR survey image of the galaxy [\ks], H-band SINFONI continuum image [Cont], SINFONI \halpha flux image [\halpha], signal-to-noise map for every spaxel [SNR], observed radial velocity map [Rot$_{obs}$], residual rotational velocity map  [Rot$_{res}$] (after subtracting quadratically the rotational velocity model from Figure \ref{fig:montage_models1} in this case, otherwise Figure \ref{fig:montage_models2}), observed velocity dispersion map [$\sigma_{obs}$] and residual velocity dispersion map [$\sigma_{res}$] (after subtracting quadratically the velocity dispersion model from Figure \ref{fig:montage_models1} in this case, otherwise Figure \ref{fig:montage_models2}). In our maps, north is up and east is left. \halpha contours are overlapped in all the maps (with decrements of 10\% in flux between adjacent contours) in order to facilitate the reader to know which spaxels belong to the galaxies. The axes show sizes both in kpc and arcsec. A cross appears in the place we have set our galaxy center (see beginning of Section \ref{sec:galaxies} for an explanation) and the seeing is represented in the lower left part of the SINFONI maps by a circle whose size is its FHWM. \newline
POWIR1 -- Disk-like galaxy. \textit{Comments:} Clear \halpha line in all the spaxels above the signal-to-noise threshold, velocity dispersion enhancement due to minor merging.}
\label{fig:montage_powir1}

\end{figure*}

The maps belonging to this galaxy are shown in Figure \ref{fig:montage_powir1}.
This galaxy, albeit a clear detection, is a very compact system both in the \ks band and in the \halpha image. The explanation for observing only few spaxels is its inclination, which is the highest of the
 sample. When looking at the continuum image other features appear. There is a flux stripe in the right side which is spurious as it extends from the left to the right part of the detector, having no
 counterpart in both \ks and \halpha maps. However, we notice two blobs which seem to be real. They have an angular size comparable with the seeing of this observation, which reassures us of their
 detections. The fainter one, near the northern edge of the primary galaxy, may help us understand why this galaxy shows large velocity dispersion values close to the blob, as this may signal a minor
 merging event. Conversely, the brightest spot on the south west of the continuum image of the galaxy is not associated with any \halpha emission, and it does not cause any significant distortion of the
 main galaxy. Kinematic models show a regular rotational gradient and a fairly high ($\sim70$ km/s) $1/$error$^{2}$ velocity dispersion. Although it is tentative to identify this galaxy as a merger, we
 prefer to classify it as very inclined disk galaxy, because of its large rotational velocity and ordered velocity field without any substantial disruption.

\subsection{POWIR2}
\label{subsec:powir2}
\begin{figure*}
\includegraphics[angle=0,width=0.70\linewidth]{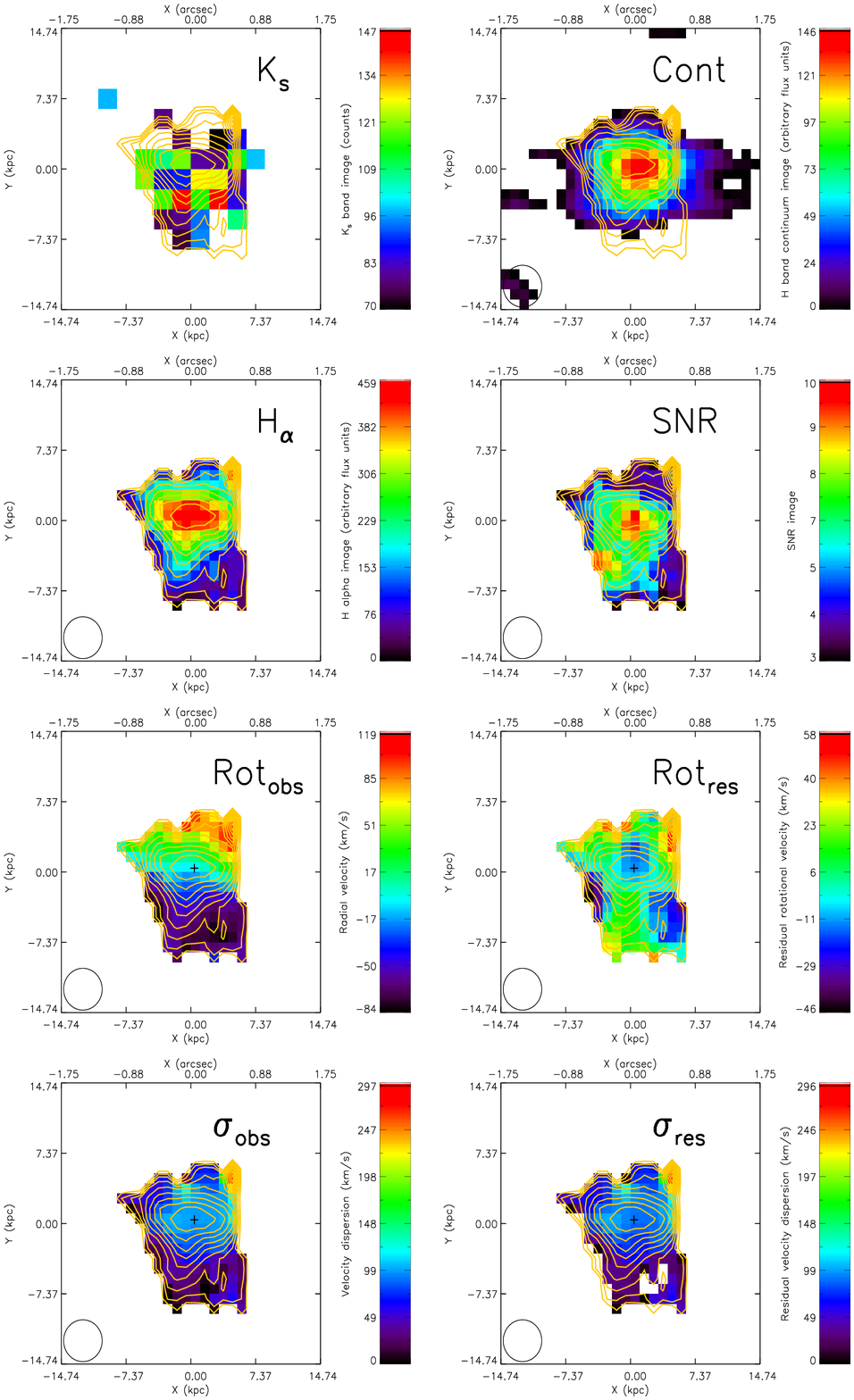}

\caption{POWIR2 -- Disk-like galaxy. \textit{Comments:} Very extended \halpha emission, which is very clearly detected. There is an elongation and a signal-to-noise enhancement in the southern part, arguably because of a merging episode. The group of spaxels in the top right display high velocity dispersion are an artefact, but not the high values in its center. This latter feature points towards the development of a spheroidal component.}
\label{fig:montage_powir2}

\end{figure*}

The maps belonging to this galaxy are shown in Figure \ref{fig:montage_powir2}.
The \halpha flux map covers in this case the whole \ks image of the galaxy. The continuum center and \halpha center are well aligned. The \halpha map peaks in the center, and there is also a very bright
 group of spaxels on the north west part of the galaxy. This sharp feature is most probably caused by a cosmic ray as it has no counterpart both in the continuum and SNR images. The galaxy looks
 asymmetrical in its southern part, where we found an enhancement of the SNR as well. We are perhaps witnessing an accretion of a minor object. But what we can state is that this galaxy looks like a
 relaxed and ordered system. The \halpha line lies in a spectral region far away from any sky line and thus the [NII] line is clearly identified, giving a ratio of $[NII]/$\halpha$ = 0.49$. This 1:2 ratio
 between \halpha and [NII] is preserved within the external spaxels, and may indicate the presence of an AGN. We consider this system as an early disk-like galaxy or spheroid with a disk, as its large
 velocity dispersion and low $V_{max}/\sigma$ ratio reveal. Nevertheless, the interpretation of a disk galaxy is favoured at the light of the strong velocity gradient.

\subsection{POWIR3}
\label{subsec:powir3}
\begin{figure*}
\includegraphics[angle=0,width=0.70\linewidth]{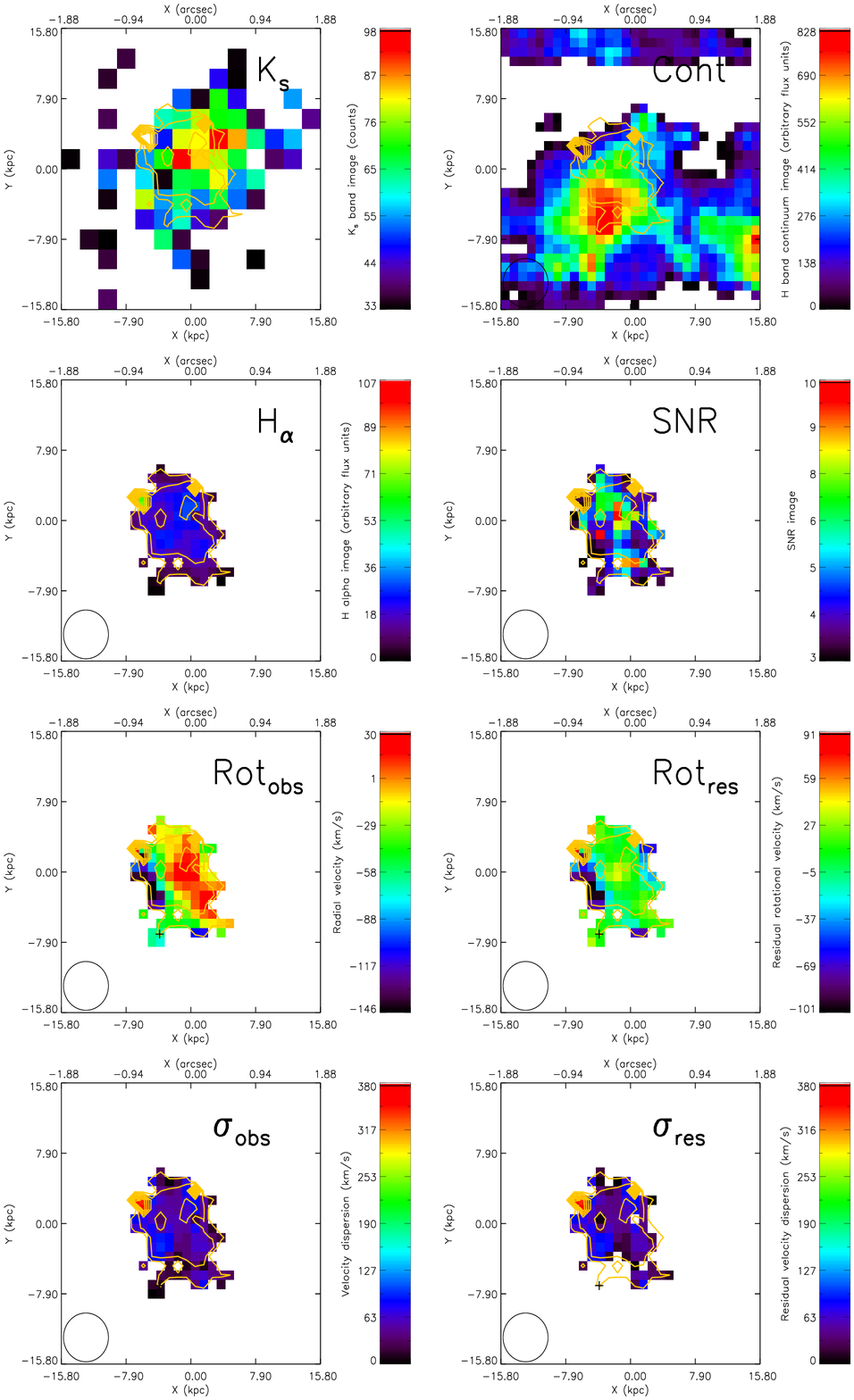}

\caption{POWIR3 -- Merging/Interacting galaxy. \textit{Comments:} \ks and continuum images are vaguely related but not with the \halpha map. Most probably this is an ongoing merger, and this would explain its elongated shape. Rotation is found but it shows complex kinematics, as expected given its interacting nature.}
\label{fig:montage_powir3}

\end{figure*}

The maps belonging to this galaxy are shown in Figure \ref{fig:montage_powir3}.
Both \ks and continuum maps show an elongated structure with a diagonal shape from east to west. \halpha countours do not exactly overlap with the galaxy continuum, and none of the brightest \halpha spots
 coincide with it. The \halpha center and the continuum center are located in different places. Regarding the continuum, we rely on its location, but not on its exact shape as it may be affected by the
 aforementioned detector problems. This is a clear case of a disturbed object, but it is remarkable that even in this case, the rotational field is quite clear. The lack of any neighbouring galaxy and
 stretched shape are evidence for an ongoing merger. With the data we have we cannot add anything to this discussion. It is important to stress that this is one of the galaxies which was observed for half
 of the integration time, and for this reason it does not have a  well-defined shape. Taking all the available probes into account we define it as disturbed/merging galaxy.

\subsection{POWIR4}
\label{subsec:powir4}
\begin{figure*}
\includegraphics[angle=0,width=0.70\linewidth]{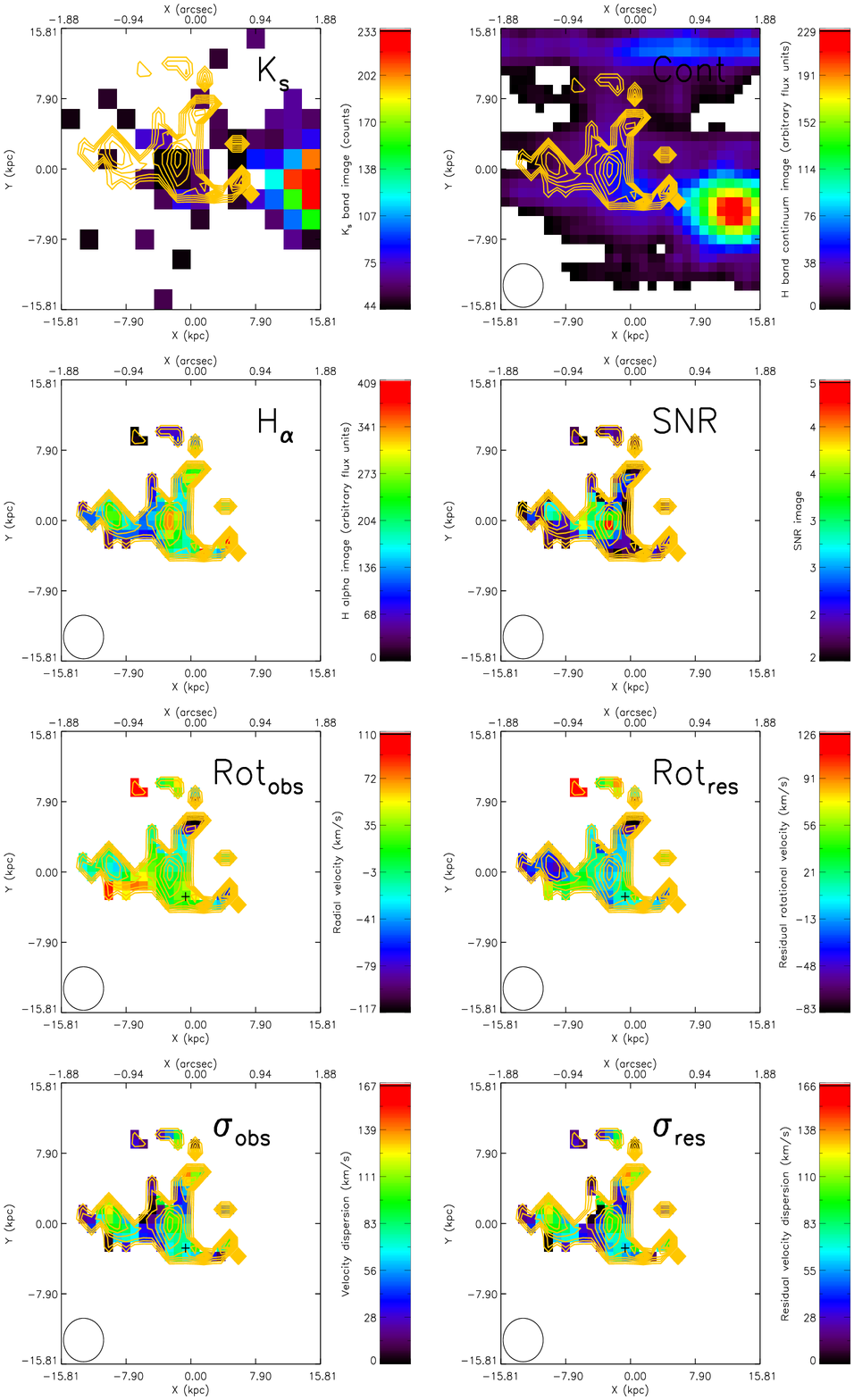}

\caption{POWIR4 -- Merging/interacting galaxy. \textit{Comments:} Both \ks and continuum maps show a bright object located at the right side of the \halpha detection. We interpret this as the massive galaxy (which is devoid of \halpha emission) is interacting with another object, whose gas is being stripped or very perturbed. For the \halpha detection, which is the non-massive galaxy, non-neglectable values of rotation and velocity dispersion are retrieved.}
\label{fig:montage_powir4}

\end{figure*}

The maps belonging to this galaxy are shown in Figure \ref{fig:montage_powir4}.
If one looks either at the \ks band map or at the continuum map a bright galaxy appears, and by its eastern side a very elongated arc-shaped feature of \halpha, which has some very weak continuum as well.
 In this case, it seems that the emission comes from a minor object whose gas has been stripped or conversely a fan of stars coming from the main object. The Palomar image also shows this feature,
 indicating that the merging interpretation is favoured. Our kinematics are thus inferred for the \halpha visible object which was not the primary target of our observations. We find a rotational velocity
 field, but it is not large (perhaps due to its non-massive nature) and a comparatively big velocity dispersion consistent with its interacting nature. Its morphology is clumpy, with at least two clumps
 being visible. We catalog it as disturbed/merging galaxy.

\subsection{POWIR5}
\label{subsec:powir5}
\begin{figure*}
\includegraphics[angle=0,width=0.70\linewidth]{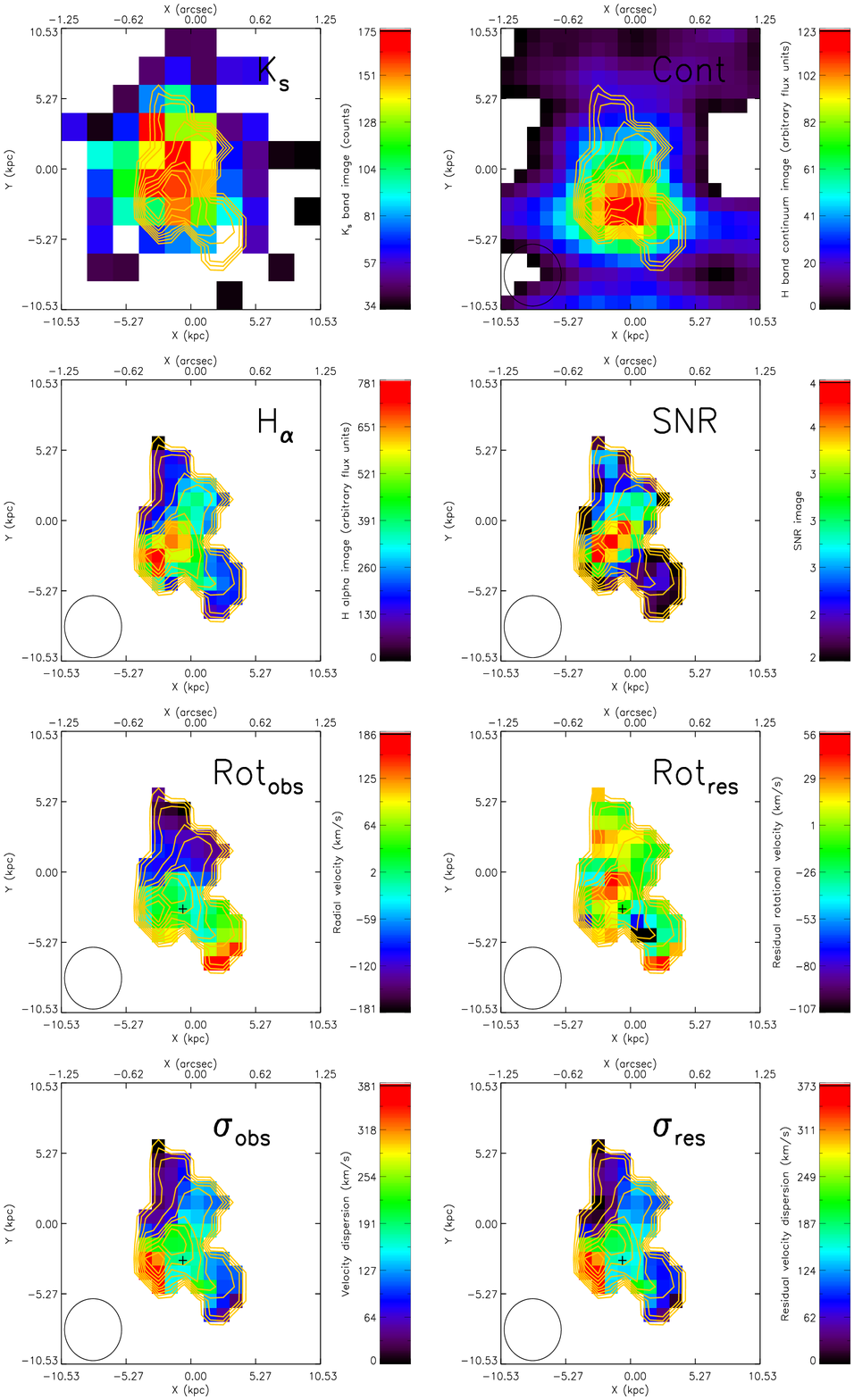}

\caption{POWIR5 -- Perturbed rotator. \textit{Comments:} \ks, continuum and \halpha overlap in the same location, showing a galaxy slightly elongated in its top part. This galaxy shows high radial velocity values. Velocity dispersion might be affected by a close sky line, increasing slightly its actual value.}
\label{fig:montage_powir5}

\end{figure*}

The maps belonging to this galaxy are shown in Figure \ref{fig:montage_powir5}.
By looking at the flux images, this galaxy appears as a blob with a tail in its upper part. There might be companions according to the POWIR/DEEP2 imaging, but we cannot state this with certainty. Its
 \halpha emission extends over the \ks image and the continuum image. The maximum rotational velocity is very high, as is the velocity dispersion, and the result is that the dynamical mass is the second
 largest in our sample. The velocity dispersion must be artificially increased by a contribution from a sky line in the redder part of the spectrum. The pipeline is meant to remove OH sky lines, but there
 are sometimes (as in this case) residuals. When the \halpha line and the sky line are close to each other it becomes very hard for our IDL programs to disentangle them, increasing somewhat the final
 result, although not significantly, as our routines were able to resolve the gradient in the radial map. We find $[NII]/$\halpha$ = 0.60$, which is puzzling as it is a large ratio and we note that the
 [NII] is detectable all over the galaxy spaxels, and not just concentrated in the center, as we would expect for AGN emission. The properties of this massive galaxy cannot be perfectly explained by a 
disk-like object and for this reason we classify it as perturbed galaxy.

\subsection{POWIR6}
\label{subsec:powir6}
\begin{figure*}
\includegraphics[angle=0,width=0.70\linewidth]{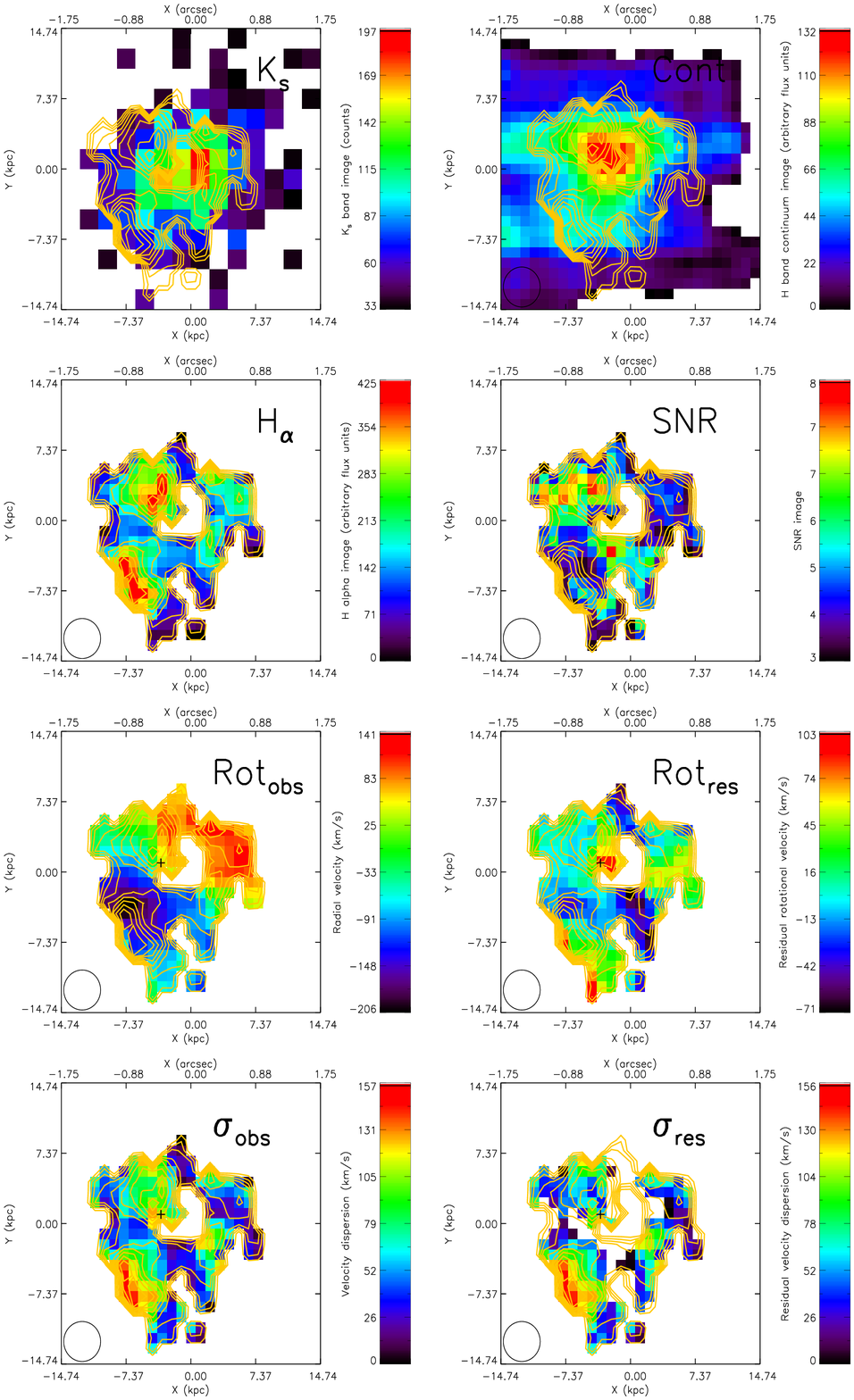}

\caption{POWIR6 -- Disk-like galaxy. \textit{Comments:} This ring pattern in the \halpha image was observed before in low redshift early-type spiral galaxies (see Epinat et al. 2010). We identify four different clumps in \halpha and the velocity dispersion map. }
\label{fig:montage_powir6}

\end{figure*}

The maps belonging to this galaxy are shown in Figure \ref{fig:montage_powir6}.
This is the brightest galaxy in our POWIR imaging. The \halpha appearance of the galaxy is different from either the continuum or the \ks band image. Remarkably, we do not detect \halpha emission in the
 central area, appearing as a hole in the \halpha flux map and four knots or clumps surrounding it. This hole overlaps with the center of the \ks and continuum images. Similar cases are seen in Epinat et
 al. (2010), for example after redshifting the local galaxy UGC04820 where its ring nature produces the effects observed in our flux maps. This is typically found in early-type spiral galaxies. The clumps
 are also conspicuous in the velocity dispersion map. Both theoretical expectations and other recent work agree with a likely clumpy phase in galaxy formation at high-z (F\"{o}rster-Schreiber et al. 2011b
 and references therein). We identify four of them in the \halpha map, which match the velocity dispersion enhancements. Another piece of evidence for a star-forming disk is that it is strongly rotationally supported ($V_{max}/\sigma = 4.88$ and $V_{max} = 214$ km s$^{-1}$). Hence, we classify it as a disk galaxy.

\subsection{POWIR7}
\label{subsec:powir7}
\begin{figure*}
\includegraphics[angle=0,width=0.70\linewidth]{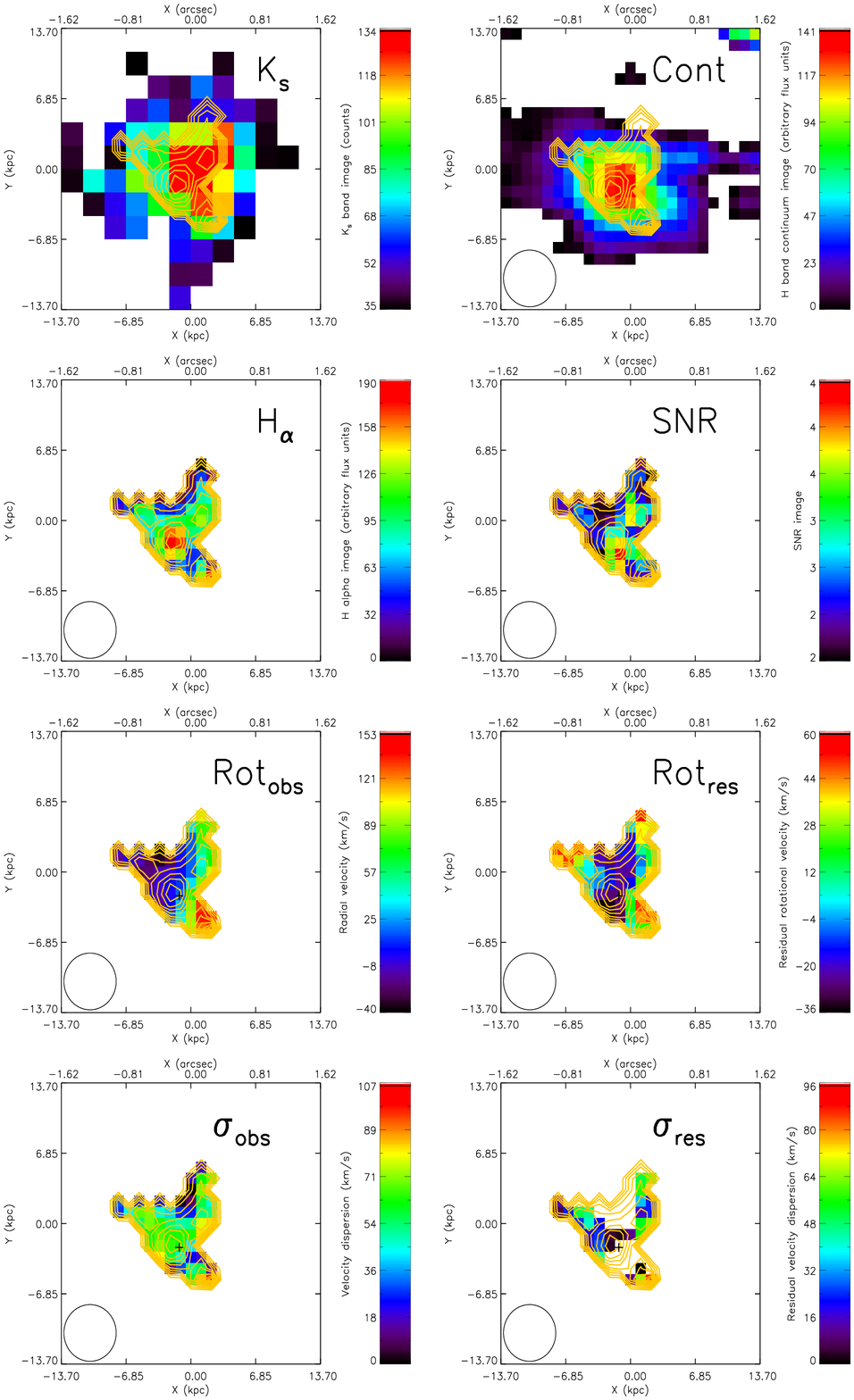}

\caption{POWIR7 -- Perturbed rotator. \textit{Comments:} \ks, \halpha and continuum maps overlap well. This galaxy was not observed half of the integration time but was set in the border of the detector, which explains its low signal-to-noise detection. Slightly disturbed morphology, with a possible clumpy structure.}
\label{fig:montage_powir7}

\end{figure*}

The maps belonging to this galaxy are shown in Figure \ref{fig:montage_powir7}.
In this case, one of the images of the galaxy fell in the borders of the SINFONI detector. This is the reason for its low SNR in all its spaxels. However, \ks POWIR/DEEP imaging, \halpha and continuum maps
 overlap well, and show a slightly distorted and clumpy galaxy. Rotation seems to play an important role in the support of this galaxy (with $V_{max}/\sigma = 6.88$), but the low SNR prevents us from
 drawing any further conclusions. Because of its irregular shape, we consider this is a perturbed galaxy.

\subsection{POWIR8}
\label{subsec:powir8}
\begin{figure*}
\includegraphics[angle=0,width=0.70\linewidth]{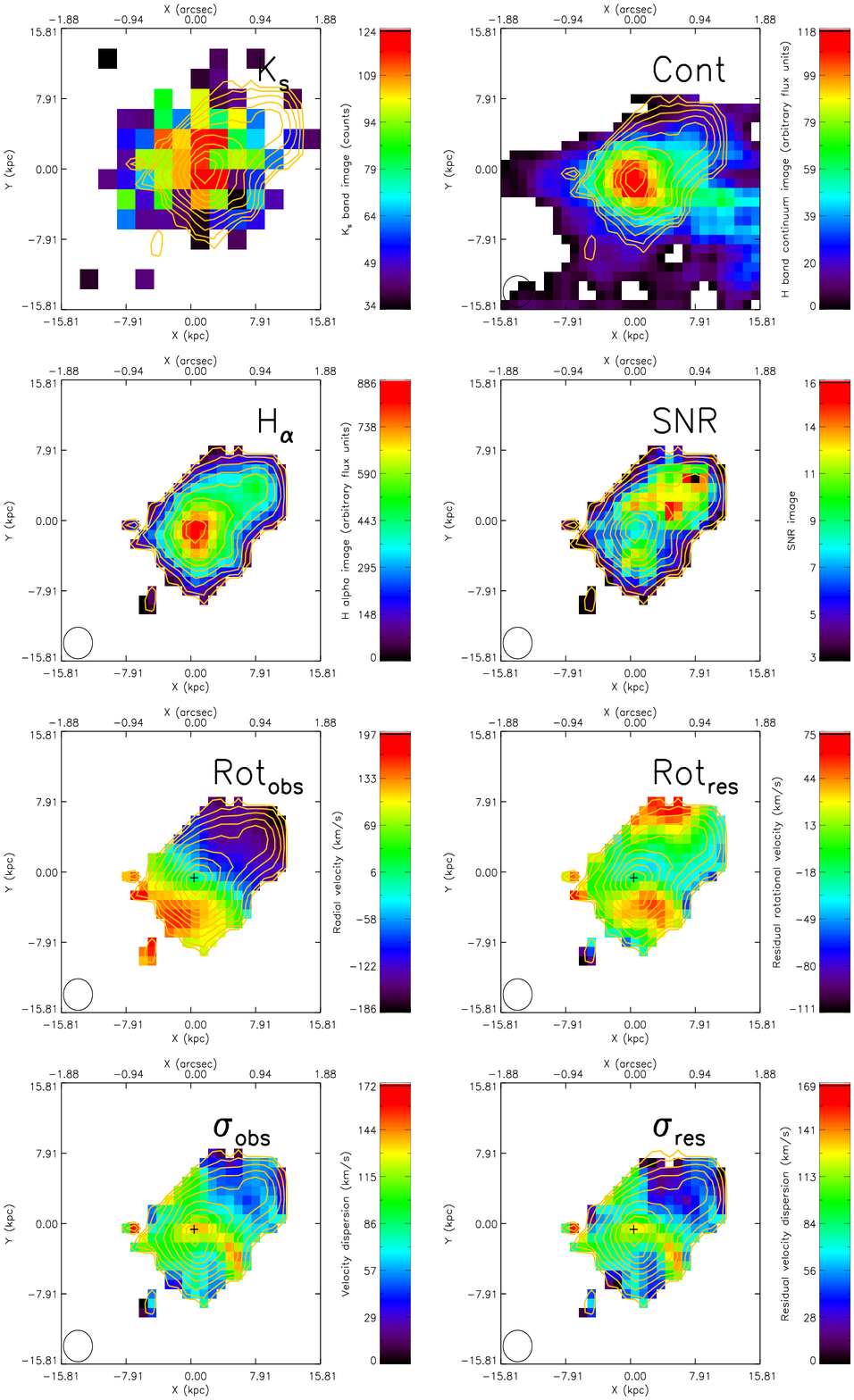}

\caption{POWIR8 -- Disk-like galaxy. \textit{Comments:} Clear and extended disk in all the images. High radial velocity values, with a large velocity dispersion in its center, which is a hint of a bulge component.}
\label{fig:montage_powir8}

\end{figure*}

The maps belonging to this galaxy are shown in Figure \ref{fig:montage_powir8}.
Our best seeing (0.42 arcsec) observations are for this system where we retrieve a very clear disk. A bulge component in the center may be present, as the central region is very bright in both \halpha and
 \ks imaging, and displays a large velocity dispersion of $\sigma\sim130$ km/s. This agrees with the renditions of the anisotropy plot (which will be explained in the Section \ref{subsec:rot_vs_sigma}), as
 this system is found near a locus close to the local fast rotators. However, an alternative explanation could be that our models were not able to resolve completely the rotation in the inner part of the
 galaxy, in which case we would have a ring distribution of gas as in POWIR6. The rotational support is very strong ($V_{max}/\sigma = 4.60$). Its dynamical mass is also very large 
($6.63\times10^{11} M_{\odot}$) and reliably measured due to its high rotational velocity gradient, the identification of which was helped by the excellent seeing of this object's observation. We classify
 this galaxy as a rotating disk.

\subsection{POWIR10}
\label{subsec:powir10}
\begin{figure*}
\includegraphics[angle=0,width=0.70\linewidth]{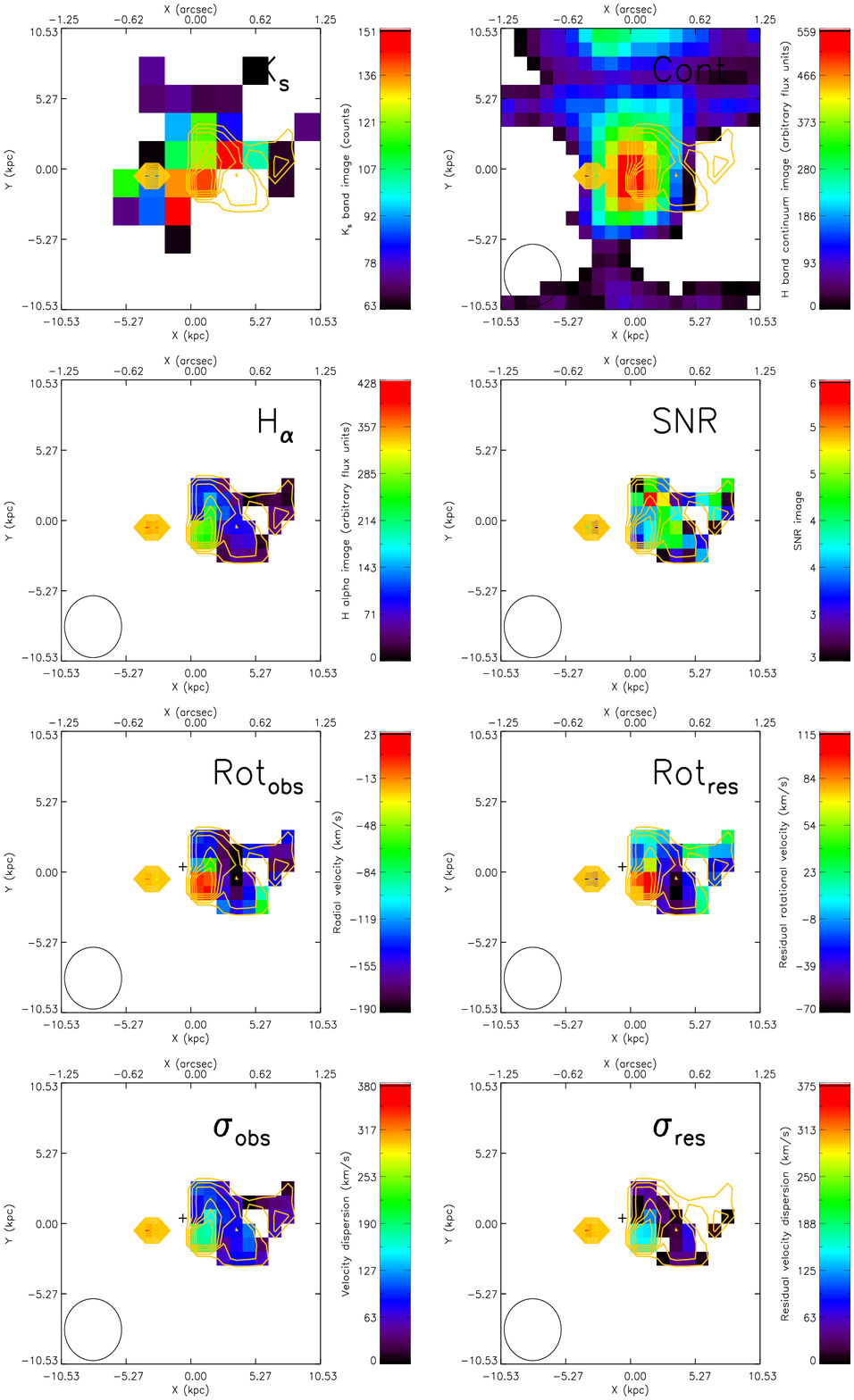}

\caption{POWIR10 -- Merging/interacting object. \textit{Comments:} Similar case as POWIR4. Here we can see the main object, which has another blob in its northern part. As most of the continuum and \ks signal come from this main object, we identify it as the massive galaxy. It show high [NII] values in its centers, indicative of its AGN nature. The \halpha detections at its sides might be related with gas outflows. The kinematics are messy, we cannot infer anything conclusive.}
\label{fig:montage_powir10_agn}

\end{figure*}

\begin{figure*}
\includegraphics[angle=0,width=0.70\linewidth]{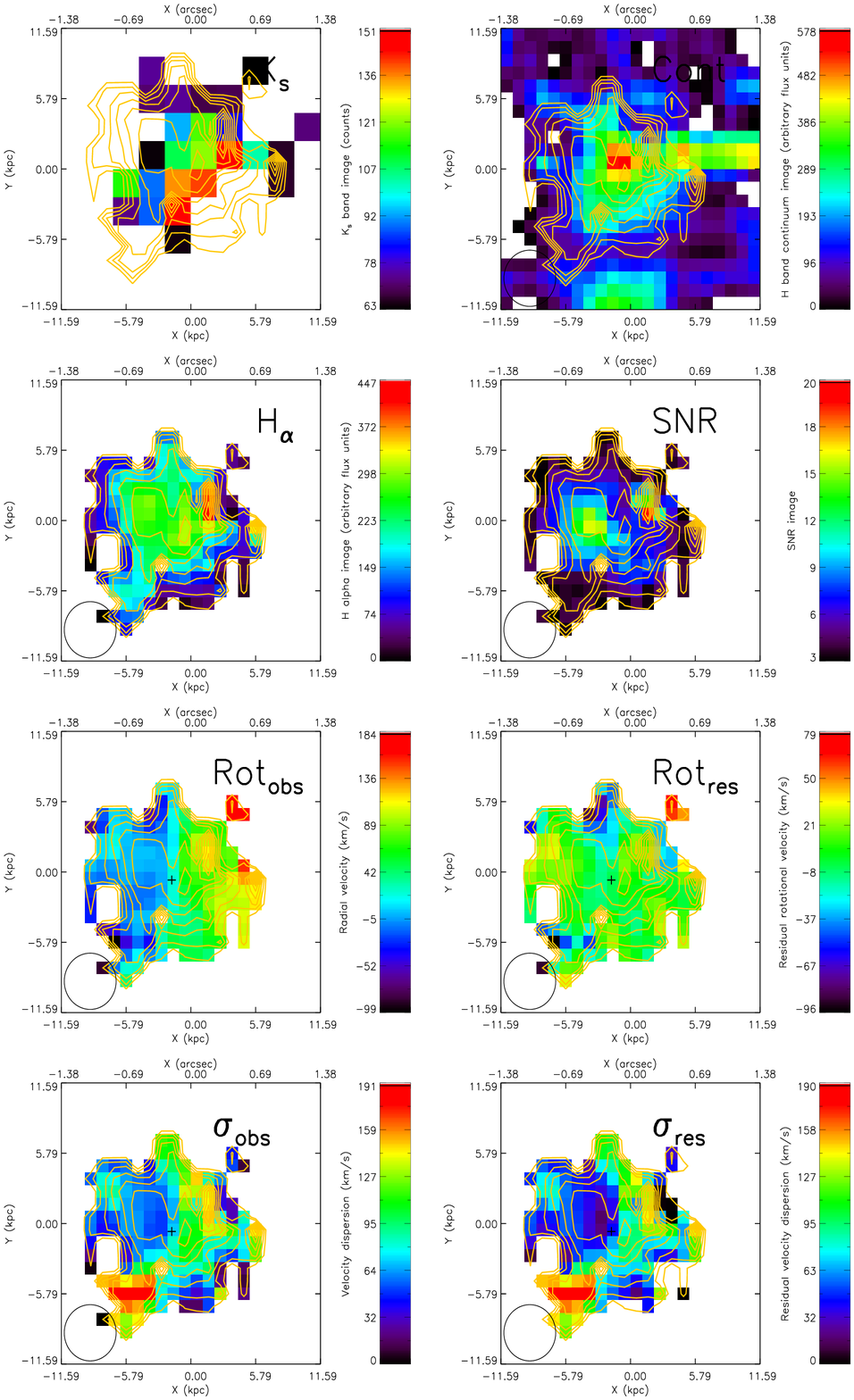}

\caption{POWIR10 -- Merging/interacting object. \textit{Comments:} This is the galaxy which was at the north of the previous Figure \ref{fig:montage_powir10_agn} object. It is more extended, at least in the \halpha map. Its radial and velocity dispersion values are not very high, suggesting it is not as massive as its partner galaxy. There is a velocity dispersion enhancement in the southern part, where the two galaxies are connected/interacting.}
\label{fig:montage_powir10_no_agn}

\end{figure*}

The maps belonging to this galaxy are shown in Figures \ref{fig:montage_powir10_agn} and \ref{fig:montage_powir10_no_agn}.
We present two different set of figures for this galaxy. Our \ks and continuum images show two galaxies. Most of the flux in these two maps comes from the object in the southern part of the image (Figure
 \ref{fig:montage_powir10_agn}), although \halpha emission is mainly found in the galaxy on the north (Figure \ref{fig:montage_powir10_no_agn}). This system has some similarities with POWIR4, specifically
 in that most of the \halpha emission does not come from the position where we had expected to find the massive galaxy. For this reason, the \halpha kinematic parameters listed on the tables as POWIR10 are
 associated with the object identified as the non-massive galaxy, as kinematics could only be reliably retrieve from it. As a matter of fact, we found strong emission at the [NII] wavelength for the
 central spaxels of the southern object. This makes us suspect that, as it is nearly a point source, this system harbours an AGN on its center, and the \halpha emission at its sides might be related with
 shocked gas by outflows of material coming out from it. Regarding the non-massive galaxy within our imaging, we are able to detect a hint of continuum and very clear kinematics. Due to the low values in
 the kinematic velocities for this second galaxy we conclude that the photometric mass is only derived for the AGN. For the total system, due to its distortion/clumpiness, the velocity dispersions
 enhancement in between the two objects and of course since \halpha is detected at the same wavelength for the two subsystems, we classify this object as a merger. 

\subsection{POWIR11}
\label{subsec:powir11}
\begin{figure*}
\includegraphics[angle=0,width=0.70\linewidth]{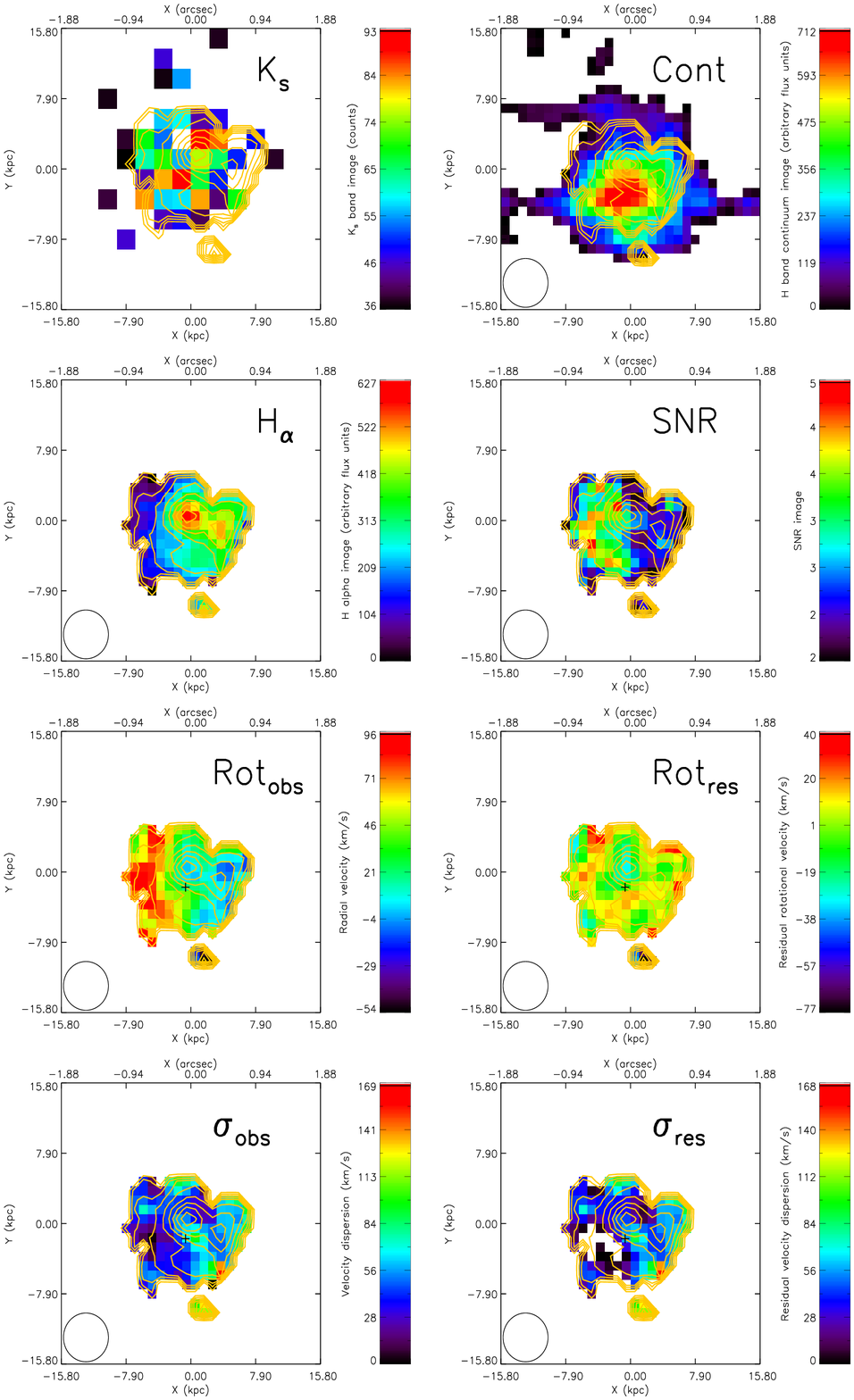}

\caption{POWIR11 -- Disk-like object. \textit{Comments:} Close agreement between \ks, continuum and \halpha light distributions. This galaxy could only be observed for half of the nominal integration time, retrieving a low signal-to-noise per spaxel. This is the galaxy with the lowest inclination in our sample. Clear velocity gradient as \halpha is clearly detected in all the coloured spaxels.}
\label{fig:montage_powir11}

\end{figure*}

The maps belonging to this galaxy are shown in Figure \ref{fig:montage_powir11}.
This is a clearly detected galaxy which was only observed half of the time as the other systems. However, \halpha emission show an extended galaxy, in agreement with the \ks and continuum images. It is
 most probably close to being face-on, as its inclination is the smallest (30$^\circ$) in our sample. The SNR decreases in its western side because of the presence of a sky line at the same side of the
 spectrum where the \halpha line is found. Nevertheless, the \halpha line is clearly distinguished in all the spaxels up to the faint outskirts, making the rotational velocity gradient easy to detect. This
 was the reason behind setting $S/N =2$ as a threshold for the galaxy maps. Overall, this galaxy presents a disturbed discoidal shape and a couple of bright \halpha spots or clumps in its central part. The
 absence of any merger signature makes us conclude that this is a disk, although rather perturbed or turbulent, as indicated by its fairly low $V_{max}/\sigma$ ratio.

\clearpage 

\clearpage


\end{document}